\documentclass[twocolumn,traditabstract,a4]{aa}
\usepackage{graphicx,amsmath}
\usepackage{epsf}
\usepackage{txfonts}
\usepackage[hyphens]{url}
\usepackage{longtable}
\usepackage{lscape}
\usepackage[hyperindex,breaklinks=true, colorlinks, citecolor=blue]{hyperref}  % for active links 
\usepackage{breakurl}
\usepackage{comment}
\usepackage{natbib}
\usepackage{upgreek}
\usepackage{float}
\usepackage[switch,pagewise]{lineno}
\usepackage{multirow}
\newcommand{\hpeter}[1]{}
\newcommand{\juan}[1]{{\color{black}#1}}
\newcommand{\commentproof}[1]{{\bf \color{green}#1}}
\renewcommand{\commentproof}[1]{} % turn off mark-up

\def\setsymbol#1#2{\expandafter\def\csname #1\endcsname{#2}}
\def\getsymbol#1{\csname #1\endcsname}

\def\Planck{\textit{Planck}}

\def\all2013resultspapers{\nocite{planck2013-p01, planck2013-p02, planck2013-p02a, planck2013-p02d, planck2013-p02b, planck2013-p03, planck2013-p03c, planck2013-p03f, planck2013-p03d, planck2013-p03e, planck2013-p01a, planck2013-p06, planck2013-p03a, planck2013-pip88, planck2013-p08, planck2013-p11, planck2013-p12, planck2013-p13, planck2013-p14, planck2013-p15, planck2013-p05b, planck2013-p17, planck2013-p09, planck2013-p09a, planck2013-p20, planck2013-p19, planck2013-pipaberration, planck2013-p05, planck2013-p05a, planck2013-pip56, planck2013-p06b, planck2013-p01a}}

\newbox\tablebox    \newdimen\tablewidth
\def\leaderfil{\leaders\hbox to 5pt{\hss.\hss}\hfil}
\def\endPlancktable{\tablewidth=\columnwidth 
    $$\hss\copy\tablebox\hss$$
    \vskip-\lastskip\vskip -2pt}

\def\tablenote#1 #2\par{\begingroup \parindent=0.8em
    \abovedisplayshortskip=0pt\belowdisplayshortskip=0pt
    \noindent
    $$\hss\vbox{\hsize\tablewidth \hangindent=\parindent \hangafter=1 \noindent
    \hbox to \parindent{$^#1$\hss}\strut#2\strut\par}\hss$$
    \endgroup}
\def\doubleline{\vskip 3pt\hrule \vskip 1.5pt \hrule \vskip 5pt}

\def\L2{\ifmmode L_2\else $L_2$\fi}

\def\DeltaT{\ifmmode \Delta T\else $\Delta T$\fi}
\def\deltat{\ifmmode \Delta t\else $\Delta t$\fi}
\def\fknee{\ifmmode f_{\rm knee}\else $f_{\rm knee}$\fi}
\def\Fmax{\ifmmode F_{\rm max}\else $F_{\rm max}$\fi}
\def\solar{\ifmmode{\rm M}_{\mathord\odot}\else${\rm M}_{\mathord\odot}$\fi}
\def\Msolar{\ifmmode{\rm M}_{\mathord\odot}\else${\rm M}_{\mathord\odot}$\fi}
\def\Lsolar{\ifmmode{\rm L}_{\mathord\odot}\else${\rm L}_{\mathord\odot}$\fi}
\def\inv{\ifmmode^{-1}\else$^{-1}$\fi}
\def\mo{\ifmmode^{-1}\else$^{-1}$\fi}
\def\sup#1{\ifmmode ^{\rm #1}\else $^{\rm #1}$\fi}
\def\expo#1{\ifmmode \times 10^{#1}\else $\times 10^{#1}$\fi}
\def\,{\thinspace}
\def\lsim{\mathrel{\raise .4ex\hbox{\rlap{$<$}\lower 1.2ex\hbox{$\sim$}}}}
\def\gsim{\mathrel{\raise .4ex\hbox{\rlap{$>$}\lower 1.2ex\hbox{$\sim$}}}}

\def\simprop{\mathrel{\raise .4ex\hbox{\rlap{$\propto$}\lower 1.2ex\hbox{$\sim$}}}}
\def\deg{\ifmmode^\circ\else$^\circ$\fi}
\def\pdeg{\ifmmode $\setbox0=\hbox{$^{\circ}$}\rlap{\hskip.11\wd0 .}$^{\circ}
          \else \setbox0=\hbox{$^{\circ}$}\rlap{\hskip.11\wd0 .}$^{\circ}$\fi}
\def\arcs{\ifmmode {^{\scriptstyle\prime\prime}}
          \else $^{\scriptstyle\prime\prime}$\fi}
\def\arcm{\ifmmode {^{\scriptstyle\prime}}
          \else $^{\scriptstyle\prime}$\fi}
\newdimen\sa  \newdimen\sb
\def\parcs{\sa=.07em \sb=.03em
     \ifmmode \hbox{\rlap{.}}^{\scriptstyle\prime\kern -\sb\prime}\hbox{\kern -\sa}
     \else \rlap{.}$^{\scriptstyle\prime\kern -\sb\prime}$\kern -\sa\fi}
\def\parcm{\sa=.08em \sb=.03em
     \ifmmode \hbox{\rlap{.}\kern\sa}^{\scriptstyle\prime}\hbox{\kern-\sb}
     \else \rlap{.}\kern\sa$^{\scriptstyle\prime}$\kern-\sb\fi}
\def\ra[#1 #2 #3.#4]{#1\sup{h}#2\sup{m}#3\sup{s}\llap.#4}
\def\dec[#1 #2 #3.#4]{#1\deg#2\arcm#3\arcs\llap.#4}
\def\deco[#1 #2 #3]{#1\deg#2\arcm#3\arcs}
\def\rra[#1 #2]{#1\sup{h}#2\sup{m}}

\def\dots{\relax\ifmmode \ldots\else $\ldots$\fi}
\def\WHzsr{\ifmmode $W\,Hz\mo\,sr\mo$\else W\,Hz\mo\,sr\mo\fi}
\def\mHz{\ifmmode $\,mHz$\else \,mHz\fi}
\def\GHz{\ifmmode $\,GHz$\else \,GHz\fi}
\def\mKs{\ifmmode $\,mK\,s$^{1/2}\else \,mK\,s$^{1/2}$\fi}
\def\muKs{\ifmmode \,\mu$K\,s$^{1/2}\else \,$\mu$K\,s$^{1/2}$\fi}
\def\muKRJs{\ifmmode \,\mu$K$_{\rm RJ}$\,s$^{1/2}\else \,$\mu$K$_{\rm RJ}$\,s$^{1/2}$\fi}
\def\muKHz{\ifmmode \,\mu$K\,Hz$^{-1/2}\else \,$\mu$K\,Hz$^{-1/2}$\fi}
\def\MJysr{\ifmmode \,$MJy\,sr\mo$\else \,MJy\,sr\mo\fi}
\def\MJysrmK{\ifmmode \,$MJy\,sr\mo$\,mK$_{\rm CMB}\mo\else \,MJy\,sr\mo\,mK$_{\rm CMB}\mo$\fi}
\def\microns{\ifmmode \,\mu$m$\else \,$\mu$m\fi}
\def\micron{\microns}
\def\muK{\ifmmode \,\mu$K$\else \,$\mu$\hbox{K}\fi}
\def\microK{\ifmmode \,\mu$K$\else \,$\mu$\hbox{K}\fi}
\def\muW{\ifmmode \,\mu$W$\else \,$\mu$\hbox{W}\fi}
\def\kms{\ifmmode $\,km\,s$^{-1}\else \,km\,s$^{-1}$\fi}
\def\kmsMpc{\ifmmode $\,\kms\,Mpc\mo$\else \,\kms\,Mpc\mo\fi}
\providecommand{\sorthelp}[1]{}

\def\BLASTPol{BLASTPol}
\def\Herschel{\textit{Herschel}}
\newcommand{\nh}{$N_{\textsc{H}}$}
\newcommand{\nhd}{N_{\textsc{H}}} % use in math mode

\newcommand{\lognh}{$\log_{10}(N_{\textsc{H}}/\mbox{cm}^{-2})$}

\newcommand{\bperp}{$\langle\hat{\vec{B}}_{\perp}\rangle$}

\newcommand{\healpix}{{\sc HEALPix}}

\providecommand{\sorthelp}[1]{}

\begin{document}
%%%%%%%%%%%%%%%%%%%%%%%%%%%%%%%%%%%%%%%%%%%%%%%%%%%%%%%%%%%%
%\linenumbers
%%%%%%%%%%%%%%%%%%%%%%%%%%%%%%%%%%%%%%%%%%%%%%%%%%%%%%%%%%%%

%\input PIP_113_Boulanger_authors_and_institutes.tex

\title{On the relation between the column density structures and the magnetic field orientation in the Vela\,C molecular complex}
\titlerunning{Relative orientation between \bperp\ and \nh\ structures towards Vela\,C}
\author{J.~D.~Soler$^{1,2}$\thanks{Corresponding author: Juan D. Soler (soler@mpia.de)},
P.~A.~R.~Ade$^{3}$,
F.~E.~Angil\`e$^{4}$,
P.~Ashton$^{5}$,
S.~J.~Benton$^{6}$,
M.~J.~Devlin$^{4}$,
B.~Dober$^{4,7}$,
L.~M.~Fissel$^{8}$,
Y.~Fukui$^{9}$,
N.~Galitzki$^{4,10}$,
N.~N.~Gandilo$^{11}$,
P.~Hennebelle$^{2}$,
J.~Klein$^{4}$,
Z.-Y.~Li$^{12}$,
A.~L.~Korotkov$^{13}$,
P.~G.~Martin$^{14}$,
T.~G.~Matthews$^{5}$,
L.~Moncelsi$^{15}$,
C.~B.~Netterfield$^{16}$,
G.~Novak$^{5}$,
E.~Pascale$^{3}$,
F.~Poidevin$^{17,18}$,
F.~P. Santos$^{5}$,
G.~Savini$^{19}$,
D.~Scott$^{20}$,
J.~A.~Shariff$^{14}$,
N.~E.~Thomas$^{21}$,
C.~E.~Tucker$^{3}$,
G.~S.~Tucker$^{13}$,
D.~Ward-Thompson$^{22}$
}

\institute{
Max-Planck-Institute for Astronomy, K\"{o}nigstuhl 17, 69117, Heidelberg, Germany
\and Laboratoire AIM, Paris-Saclay, CEA/IRFU/SAp - CNRS - Universit\'{e} Paris Diderot, 91191, Gif-sur-Yvette Cedex, France 
\and Cardiff University, School of Physics \& Astronomy, Cardiff, CF24 3AA, U.K. 
\and Department of Physics \& Astronomy, University of Pennsylvania, Philadelphia, PA, 19104, U.S.A. 
\and Center for Interdisciplinary Exploration and Research in Astrophysics (CIERA) and Department\ of Physics \& Astronomy, Northwestern University, Evanston, IL 60208, U.S.A. 
\and Department of Physics, Princeton University, Jadwin Hall, Princeton, NJ 08544, U.S.A. 
\and National Institute of Standards and Technology (NIST), Boulder, CO 80305, U.S.A.
\and National Radio Astronomy Observatory (NRAO), Charlottesville, VA 22903, U.S.A.
\and Department of Physics and Astrophysics, Nagoya University, Nagoya 464-8602, Japan
\and Center for Astrophysics and Space Sciences. University of California, San Diego, CA 92093, U.S.A.
\and Johns Hopkins University, Baltimore, MD 21218, U.S.A.
\and Department of Astronomy, University of Virginia, Charlottesville, VA 22904, U.S.A.
\and Department of Physics, Brown University, Providence, RI 02912, U.S.A.
\and CITA, University of Toronto, Toronto, ON M5S 3H8, Canada
\and California Institute of Technology, Pasadena, CA, 91125, U.S.A.
\and Department of Physics, University of Toronto, Toronto, ON M5S 1A7, Canada
\and Instituto de Astrofísica de Canarias, E-38200 La Laguna, Tenerife, Spain. 
\and Universidad de La Laguna, Dpto. Astrofísica, E-38206 La Laguna, Tenerife, Spain
\and Department of Physics \& Astronomy, University College London, London, WC1E 6BT, U.K.
\and Department of Physics \& Astronomy, University of British Columbia, Vancouver, BC V6T 1Z1, Canada
\and NASA/Goddard Space Flight Center, Greenbelt , MD 20771, U.S.A.
\and Jeremiah Horrocks Institute, University of Central Lancashire, PR1 2HE, U.K.
}
\authorrunning{J.~D.~Soler and The BLASTPol~Collaboration}

\date{Received 14 February 2017 / Accepted nn XX 201X}

\abstract{
We statistically evaluate the relative orientation between gas column density structures, inferred from \Herschel\ submillimetre observations, and the magnetic field projected on the plane of sky, inferred from polarized thermal emission of Galactic dust observed by the Balloon-borne Large-Aperture Submillimetre Telescope for Polarimetry (BLASTPol) at 250, 350, and 500\micron, towards the Vela\,C molecular complex.
First, we find very good agreement between the polarization orientations in the three wavelength-bands, suggesting that, at the considered common angular resolution of 3\parcm0 that corresponds to a physical scale of approximately 0.61\,pc, the inferred magnetic field orientation is not significantly affected by temperature or dust grain alignment effects.
Second, we find that the relative orientation between gas column density structures and the magnetic field changes progressively with increasing gas column density, from mostly parallel or having no preferred orientation at low column densities to mostly perpendicular at the highest column densities. 
This observation is in agreement with previous studies by the \Planck\ collaboration towards more nearby molecular clouds.
Finally, we find a correspondence between (a) the trends in relative orientation between the column density structures and the projected magnetic field, and (b) the shape of the column density probability distribution functions (PDFs).
In the sub-regions of Vela\,C dominated by one clear filamentary structure, or ``ridges'', where the high-column density tails of the PDFs are flatter, we find a sharp transition from preferentially parallel or having no preferred relative orientation at low column densities to preferentially perpendicular at highest column densities.
In the sub-regions of Vela\,C dominated by several filamentary structures with multiple orientations, or ``nests'', where the maximum values of the column density are smaller than in the ridge-like sub-regions and the high-column density tails of the PDFs are steeper, such a transition is also present, but it is clearly less sharp than in the ridge-like sub-regions.
Both of these results suggest that the magnetic field is dynamically important for the formation of density structures in this region.}

\maketitle
%\allearlypapers
%\tableofcontents

\clearpage
\section{Introduction}\label{section:introduction}

Magnetic fields are believed to play an important role in the formation of density structures in molecular clouds (MCs), from filaments to cores and eventually to stars \citep{crutcher2012,heiles2012}.
However, their particular role in the general picture of MC dynamics is still controversial, mostly due to the lack of direct observations.

A crucial tool for the \juan{study} of the interstellar magnetic field is the observation of aligned dust grains, either \juan{via} the polarization of background stars seen through MCs, or \juan{via} maps of the polarization of the far-infrared or submillimetre emission from dust in the cloud \citep{hiltner1949,hildebrand1988,planck2014-XIX}.
Aspherical spinning dust particles preferentially align their rotation axis with the local direction of the magnetic field, producing linearly polarized emission that is perpendicular to the magnetic field \citep{davis1951,lazarian2000,andersson2015}.
Thus, observations of the linear polarization provide an estimate of the magnetic field orientation projected on the plane of the sky and integrated along the line of sight, \bperp.

Recent observations by the \Planck\ satellite \citep{planck2014-a01} have produced the first all-sky map of the polarized emission from dust at submillimetre wavelengths, providing an unprecedented data set in terms of sensitivity, sky coverage, and statistical significance for the study of \bperp.
Over most of the sky, \cite{planck2014-XXXII} analysed the relative orientation between column density structures and \bperp, inferred from the \Planck\ 353-GHz (850\juan{-$\mu$m}) polarization observations at 15\arcmin\ resolution, revealing that most of the elongated structures (filaments or ridges) are predominantly parallel to the \bperp\ \juan{orientation} measured on the structures. This statistical trend becomes less striking for increasing column density.

Within ten nearby ($d < 450$\,pc) Gould Belt MCs, \cite{planck2015-XXXV} measured the relative orientation between the total column density structures, \nh\ inferred from the \Planck\ dust emission observations, and \bperp, inferred from the \Planck\ 353-GHz (850\juan{-$\mu$m}) polarization observations at 10\arcmin\ resolution.
They showed that the relative orientation between \nh\ and \bperp\ changes progressively with increasing \nh, from preferentially parallel or having no preferred orientation \juan{at low \nh\ }to preferentially perpendicular \juan{at the highest \nh.}

The results presented in \cite{planck2015-XXXV} correspond to a systematic \juan{analysis} of the trends described in previous studies of the relative orientation between \nh\ structures and \bperp\ inferred from starlight polarization \citep{palmeirim2013, li2013, sugitani2011}, as confirmed by the close agreement between the \bperp\ orientations inferred from the \Planck\ 353\,GHz polarization and starlight polarization observations presented in \cite{soler2016}.
Subsequent studies of the relative orientation between \nh\ structures and \bperp\ have identified similar trends to those described in \cite{planck2015-XXXV}, using \nh\ structures derived from \Herschel\ observations at 20\arcsec\ resolution and \bperp\ inferred from the \Planck\ 353\,GHz polarization observations towards the high-latitude cloud L1642 \citep{malinen2016} \juan{as well as} using \nh\ structures derived from \Herschel\ observations \juan{together with} starlight polarization \citep{cox2016}.

The physical conditions responsible for the observed change in relative orientation between \nh\ structures and \bperp\ \juan{are} related to the degree of magnetization of the cloud \citep{hennebelle2013a,soler2013}. 
\cite{soler2013} identified similar trends in relative orientation in simulations of molecular clouds where the magnetic field is at least in equipartition with turbulence, i.e., trans- or sub-Alfv\'enic turbulence. 
This numerical interpretation, which has been further studied in \cite{chen2016}, is in agreement with the classical picture of MC formation, where the molecular cloud \juan{forms} following compression of background gas, by the passage of the \juan{G}alactic spiral shock or by an expanding supernova shell, and the compressed gas cools and so flows down the magnetic field lines to form a self-gravitating mass \citep{mestel1965,mestel1984}.

In this paper, we extend the study of the relative orientation between \nh\ structures and \bperp\ by using observations of the Vela\,C molecular complex obtained during the 2012 flight of the Balloon-borne Large-Aperture Submillimetre Telescope for Polarimetry, BLASTPol \citep{pascale2012,galitzki2014}.
Towards Vela\,C, BLASTPol provides unprecedented observations of the dust polarized emission in three different wavelength-bands \juan{(}namely, 250, 350, and 500\micron\juan{)} at 2\parcm5 resolution, thus sampling spatial scales comparable to those considered in \cite{planck2015-XXXV}, but for a more distant, more massive, and more active MC.

Previous studies by the BLASTPol collaboration include \juan{an investigation} of the relation between the total gas column density \nh, the fractional polarization $p$, and the dispersion of orientation angles observed at 500\micron\ towards the Vela\,C molecular complex\juan{,} presented in \cite{fissel2016}. 
Also using the BLASTPol data, \cite{gandilo2016} present\juan{ed} a study of the variation of $p$ in the three observed wavelength-bands towards the Vela\,C region, concluding that the polarization spectrum is relatively flat and does not exhibit a pronounced minimum at $\lambda\approx 350$\micron, as suggested by previous measurements towards other MCs. 
\juan{Additionally,} \cite{santos2016} present\juan{ed} a quantitive comparison between the near-infrared (near-IR) polarization data from background starlight and the BLASTPol observations towards Vela\,C.
\juan{In this new paper, we consider for the first time the analysis of the \bperp\ orientations derived from the BLASTPol observations towards Vela\,C.}

This paper is organized as follows.
In Section~\ref{section:region}, we present the previously observed characteristics of Vela\,C.
In Section~\ref{section:data}, we introduce the \BLASTPol\ polarization observations and the \Herschel-based estimates of total gas column density.
In Section~\ref{section:analysis}, we introduce \juan{the method of using} the histogram of relative orientations for quantifying the \juan{relation} between \nh\ structures and \bperp.
In Section~\ref{section:discussion}, we discuss the results of our analysis.
Section~\ref{section:conclusions} gives our conclusions and anticipates future work.
\juan{We reserve some additional analyses to three appendices.}
Appendix~\ref{appendix:HROI} presents the HRO analysis of Vela\,C based only on the BLASTPol observations. 
Appendix~\ref{appendix:refRegions} \juan{presents a} study of the results of the HRO analysis in the BLASTPol maps obtained with the different diffuse emission subtraction methods introduced in \cite{fissel2016}.
Finally, Appendix~\ref{appendix:AquilaRegion} presents the HRO analysis of a portion of the Aquila rift based on the \nh\ \juan{values} estimated from the \Herschel\ observations and \bperp\ estimated from the \Planck\juan{-}353\,GHz polarization observations.

\begin{figure}[ht!]
\vspace{-0.1cm}
\centerline{
\includegraphics[width=0.49\textwidth,angle=0,origin=c]{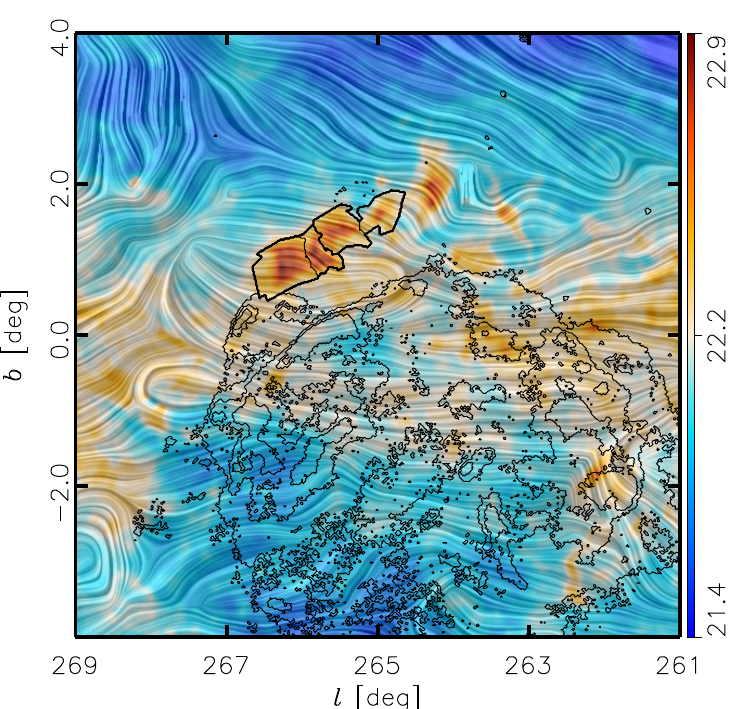}
}
\vspace{-0.1cm}
\caption{Magnetic field and column density measured by \Planck\ towards the region around the Vela\,C molecular complex. 
The colours represent the total column density, \nh, inferred from the \Planck\ observations \citep{planck2013-p06b}. 
The ``drapery'' pattern, produced using the line integral convolution method \citep[LIC,][]{cabral1993}, indicates the orientation of magnetic field lines, orthogonal to the orientation of the submillimetre polarization observed by \Planck\ at 353\GHz.
The thick black polygons correspond to the cloud sub-regions as defined in \cite{hill2011}.
The narrow black lines represent the 12.7, 42.4, and 84.9 \juan{counts per pixel contours in the X-ray energy band from 0.1 to 2.4\,keV observed with ROSAT}\citep{aschenbach1995}.
}
\label{fig:velaCzoom}
\end{figure}

% ---------------------------------------------------------------------------------------------------------------------------------------------------------------------------------------
\section{The Vela\,C region }\label{section:region}

\begin{figure}[ht!]
\centerline{
\includegraphics[width=0.48\textwidth,angle=0,origin=c]{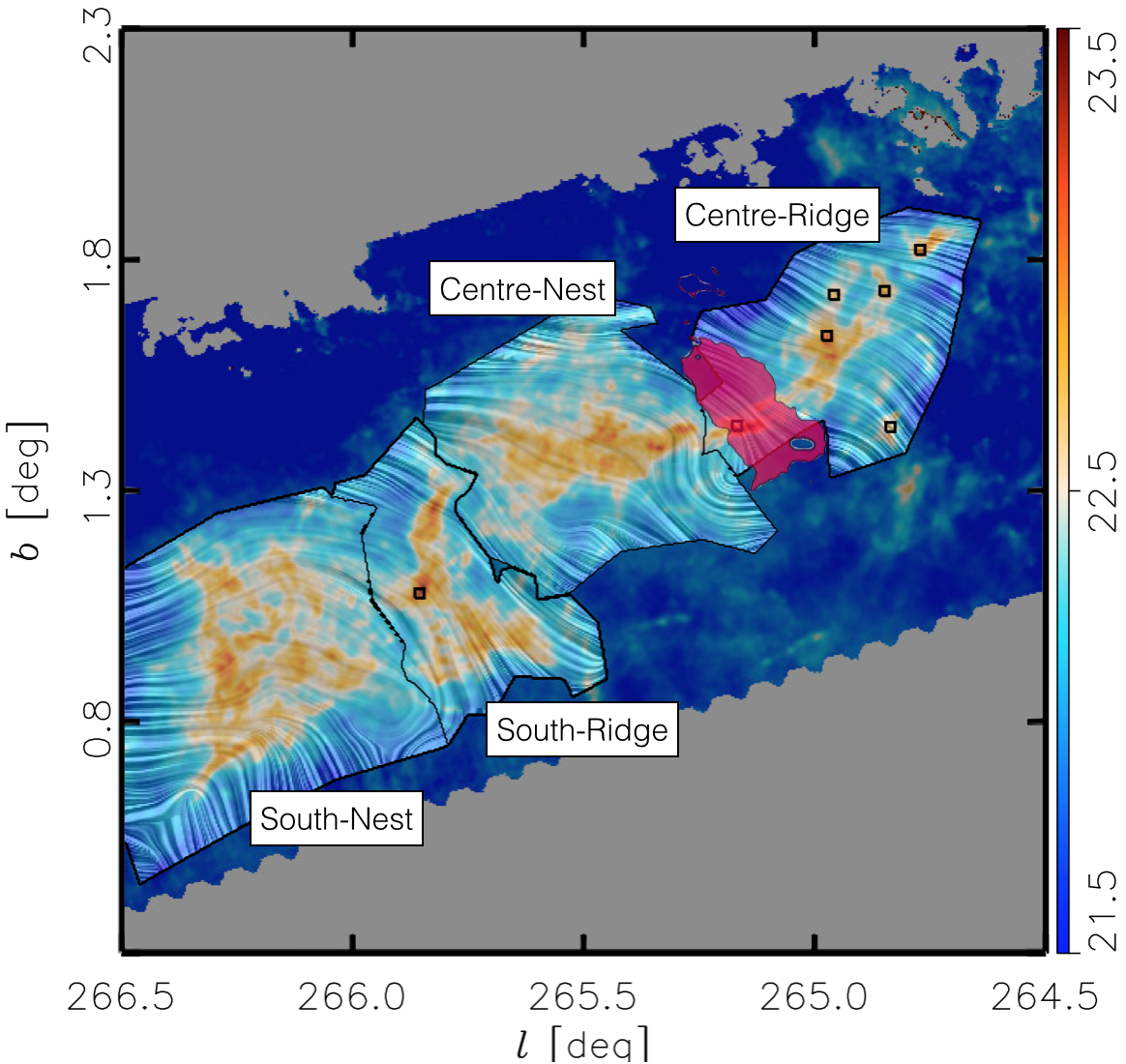}
}
\vspace{-0.1cm}
\caption{Magnetic field and total intensity measured by BLASTPol towards the analy\juan{s}ed sub-regions in the Vela\,C molecular complex. 
The colours represent \nh, the total gas column density inferred from the \Herschel\ observations. 
The ``drapery'' pattern, produced using the line integral convolution method \citep[LIC,][]{cabral1993}, indicates the orientation of magnetic field lines, orthogonal to the orientation of the submillimetre polarization observed by BLASTPol at 500\micron.
The black squares show the position\juan{s} of the dense cores \juan{with} $M>8$\,M$_{\odot}$\juan{, the mass threshold considered in \cite{hill2011},} in the catalogue presented in \cite{giannini2012}.
The \juan{large} black polygons correspond to \juan{four of the five} cloud sub-regions defined in \cite{hill2011}.
\juan{The area in magenta corresponds to the region around RCW\,36, where the dust temperature, derived from the \Herschel\ observations, is larger than 20\,K.}
}
\label{fig:VelaClumps}
\end{figure}

\juan{Figure~\ref{fig:velaCzoom} shows dust column density and magnetic field observations toward the Vela Molecular Ridge (VMR), a collection of molecular clouds lying in the Galactic plane at distances ranging from approximately 700\,pc to 2\,kpc \citep{murphy1991,liseau1992}.  
The total molecular mass of the VMR, including four distinct cloud components labeled as Vela A, B, C, and D, amounts to about $5\times10^{5}$\,M$_{\odot}$ of gas \citep{may1988,yamaguchi1999}.  
\cite{netterfield2009} presented observations of Vela\,C in dust emission at 250, 350, and 500 microns, obtained using the Balloon-borne Large Aperture Submillimetre Telescope \cite[BLAST,][]{pascale2008}.
That work confirmed that there are large numbers of objects in the early stages of star formation scattered throughout Vela\,C, along with a well-known bright compact \ion{H}{II} region, RCW\,36 \citep{baba2004}.  
\cite{hill2011} mapped the dust emission from Vela\,C using multiple Herschel wavelength bands, and identified five sub-regions with distinct characteristics which they named as North, Centre-Ridge, Centre-Nest, South-Ridge, and South-Nest. 
The last four of these were observed by BLAST-Pol and are indicated in Fig.~\ref{fig:velaCzoom} and Fig.~\ref{fig:VelaClumps}.} 

\juan{Vela\,C has long been suspected to be a rare example of a nearby \citep[$d\approx700$\,pc;][]{liseau1992} and massive \citep[$M\approx10^{5}$\,M$_{\odot}$;][]{yamaguchi1999} molecular cloud at an early evolutionary stage.
This conclusion stems from the observation that despite its relatively high mass, Vela\,C is characterized by extended regions of low temperature \citep{netterfield2009} and has produced only one or two late-type O-stars \citep[in RCW\,36;][]{baba2004}.}

% -------------------------------------------------------------------------------------------------------------------------------------------------------------------------------------
\begin{figure*}[ht!]
\centerline{
\includegraphics[width=0.33\textwidth,angle=0,origin=c]{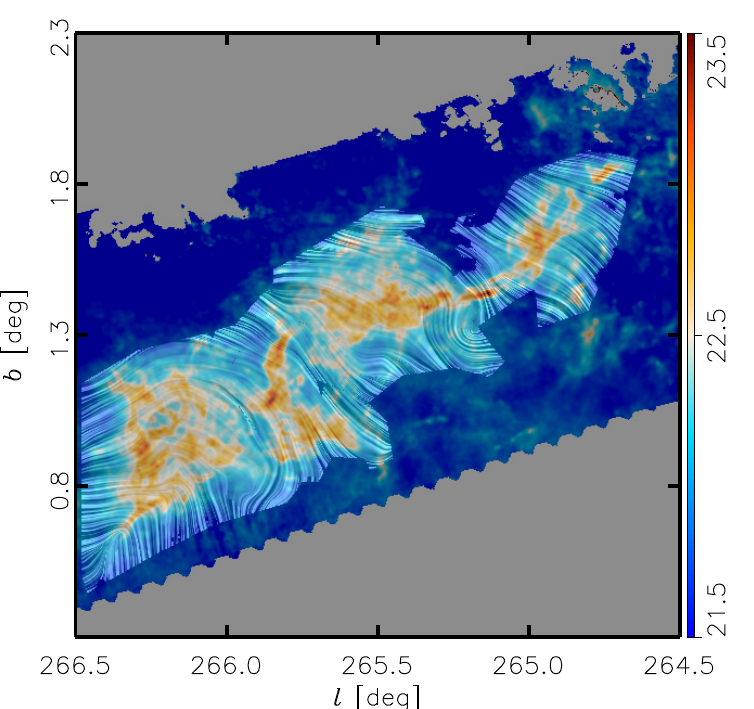}
\includegraphics[width=0.33\textwidth,angle=0,origin=c]{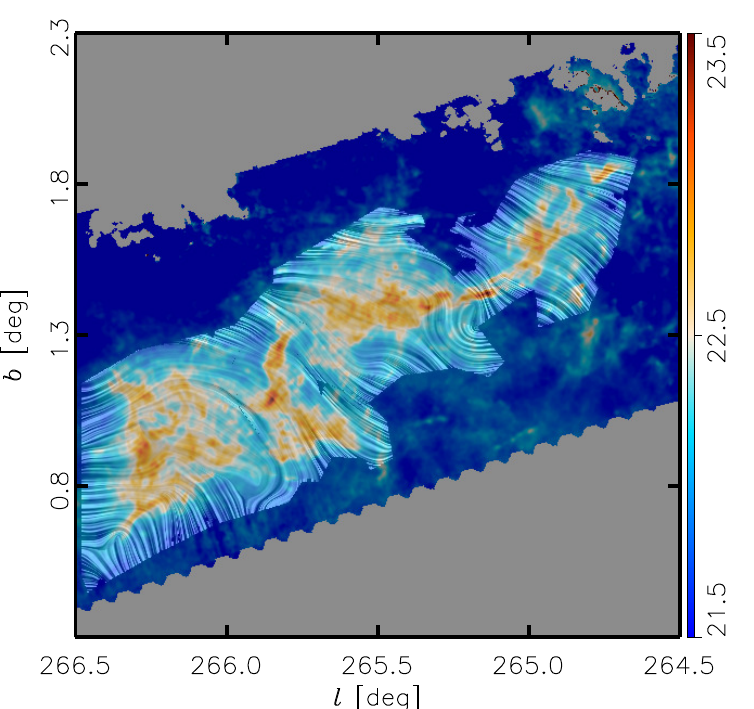}
\includegraphics[width=0.33\textwidth,angle=0,origin=c]{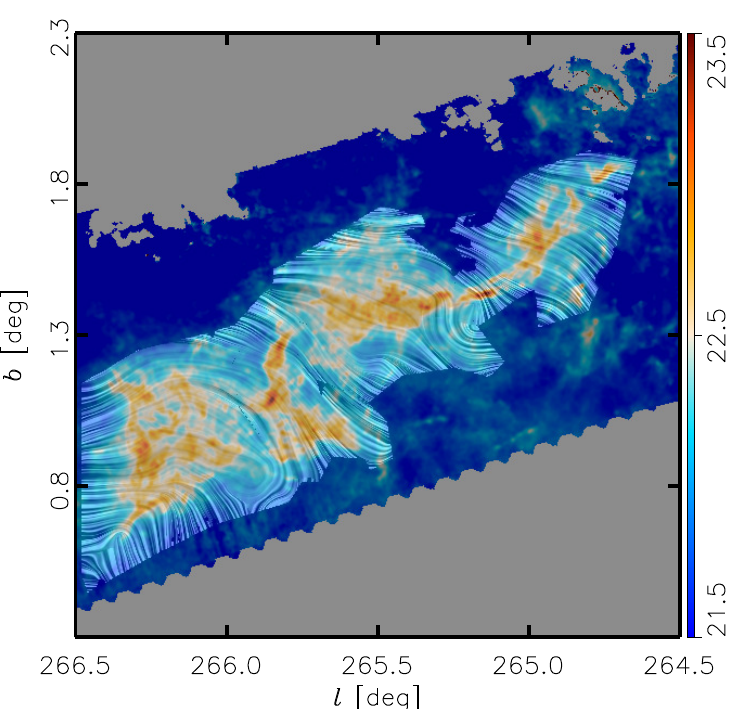}
}
\vspace{-1.0mm}
\centerline{
%\hspace{-0.01\textwidth}\includegraphics[width=0.325\textwidth,angle=0,origin=c]{HROhist250micronAll_int.eps}
%\hspace{0.01\textwidth}\includegraphics[width=0.325\textwidth,angle=0,origin=c]{HROhist350micronAll_int.eps}
%\hspace{0.01\textwidth}\includegraphics[width=0.325\textwidth,angle=0,origin=c]{HROhist500micronAll_int.eps}
\hspace{-0.01\textwidth}\includegraphics[width=0.325\textwidth,angle=0,origin=c]{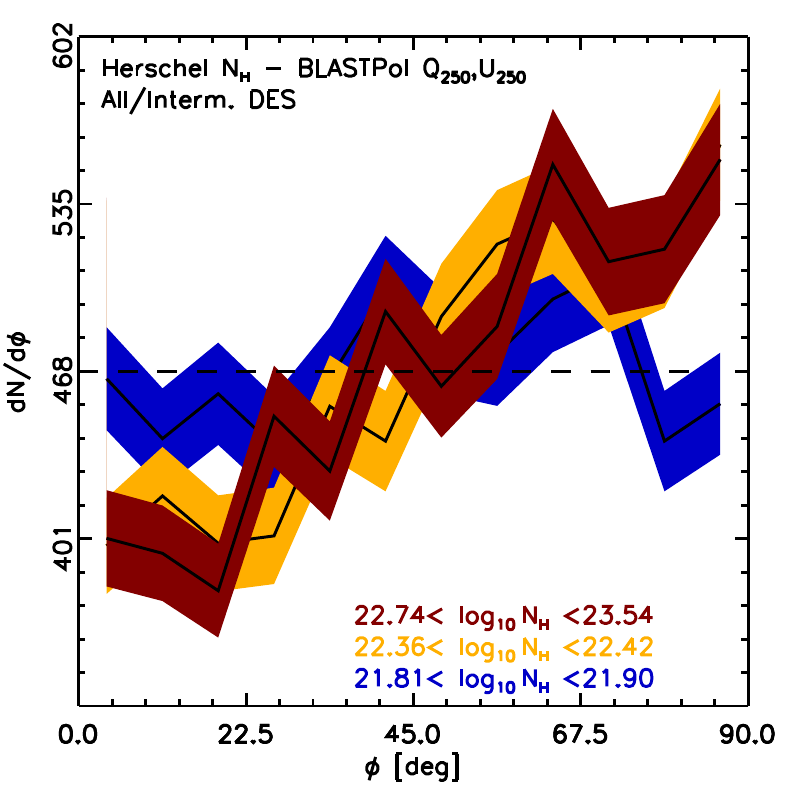}
\hspace{0.01\textwidth}\includegraphics[width=0.325\textwidth,angle=0,origin=c]{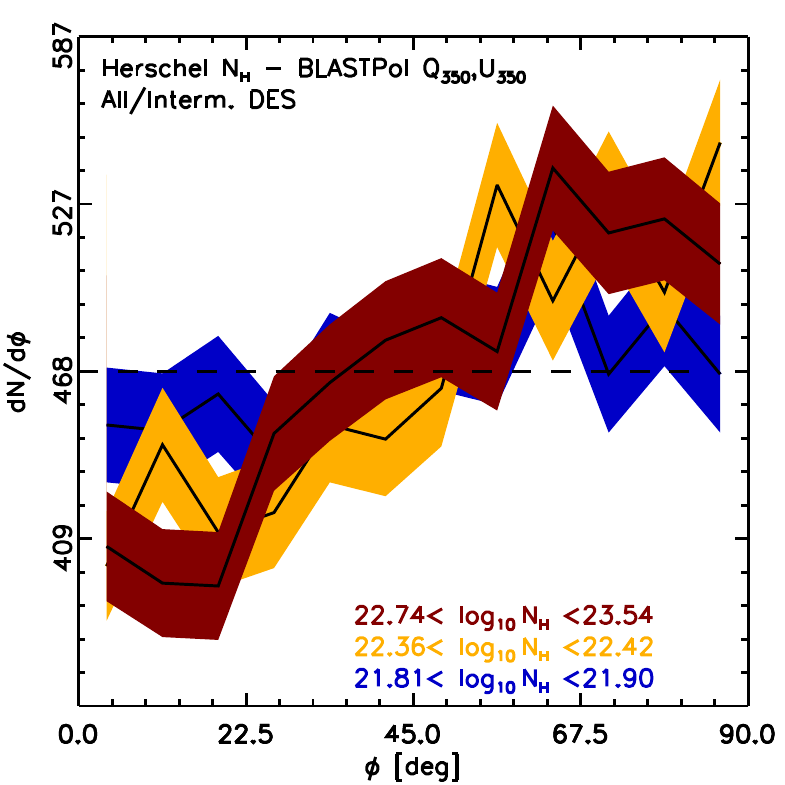}
\hspace{0.01\textwidth}\includegraphics[width=0.325\textwidth,angle=0,origin=c]{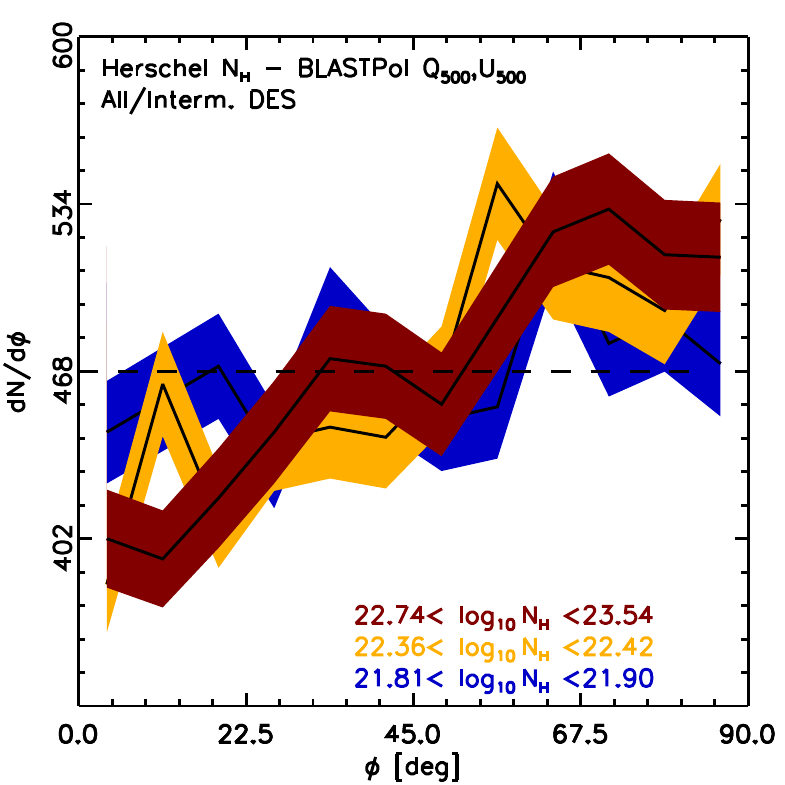}
}
\caption{\emph{Top}: Magnetic field and total intensity measured by BLASTPol towards two regions in the Vela molecular ridge. 
The colours represent \nh, the total gas column density inferred from the \Herschel\ observations.  
The ``drapery'' pattern, produced using the line integral convolution method \citep[LIC,][]{cabral1993}, indicates the orientation of magnetic field lines, orthogonal to the orientation of the submillimetre polarization observed by BLASTPol at 250 (left), 350 (centre), and 500\micron\ (right).
\emph{Bottom}: 
Histogram of the relative orientations (HRO) between the iso-\nh\ contours and the magnetic field orientation inferred from the BLASTPol observations at 250 (left), 350 (centre), and 500\micron\ (right).
The figures present the HROs for the lowest bin, an intermediate bin, and the highest \nh\ bin (blue, orange, and dark red, respectively). 
The\juan{se} bins have equal numbers of selected pixels within the \juan{indicated} \nh\ ranges.
The horizontal dashed line corresponds to the average.
The widths of the shaded areas for each histogram correspond to the 1-$\sigma$ uncertainties related to the histogram binning operation. 
Histograms peaking at 0\deg\ \juan{would} correspond to \bperp\ predominantly aligned with iso-\nh\ contours, \juan{while} histograms peaking at 90\deg\ \juan{would} correspond to \bperp\ predominantly perpendicular to iso-\nh\ contours.
}
\label{fig:HROblastpolherschel}
\end{figure*}

\section{Observations}\label{section:data}

In the present analysis we use two data sets. First, the Stokes $I$, $Q$, and $U$ observations obtained during the 2012 flight of BLASTPol. 
Second, the total column density maps derived from the \Herschel\ satellite \juan{dust continuum} observations.

\subsection{BLASTPol observations}\label{subsection:BLASTPol}

The balloon-borne submillimetre polarimeter BLASTPol and its Antarctic flights in 2010 and 2012, have been described by \cite{pascale2012}, \cite{galitzki2014}, \cite{matthews2014}, and \cite{fissel2016}.
BLASTPol used a 1.8\,m primary mirror to collect submillimetre radiation, splitting it into three wide wavelength bands ($\Delta\lambda/\lambda \approx 0.3$) centred at 250, 350, and 500\micron. 
While the telescope scanned back and forth across the target cloud, the three wavelength bands were observed simultaneously by three detector arrays operating at 300\,mK.  
The receiver optics included polarizing grids as well as an achromatic half-wave plate.
The Vela\,C observations presented here were obtained \juan{as part of} the 2012 Antarctic flight, during which the cloud was observed for 54\,hours.
The Stokes $I$, $Q$, and $U$ maps \juan{have already been} presented by \cite{fissel2016} and \cite{gandilo2016}.

\cite{fissel2016} employed three methods for subtracting the contribution that the diffuse Galactic emission makes to the measured $I$, $Q$, and $U$ maps for Vela\,C; \juan{which} referred to respectively as the ``aggressive'', ``conservative'', and ``intermediate'' methods.
The aggressive method uses two reference regions located very close to the Vela\,C cloud (one on either side of it) to estimate the levels of polarized and unpolarized emission contributed by foreground and/or background dust unassociated with the cloud.
\juan{These} contributions \juan{are then removed} from the measured $I$, $Q$, and $U$ maps.  
Because the reference regions are so close to the cloud, it is likely that they include some flux from material associated with Vela\,C.  
Thus, this method may over-correct, hence the name ``aggressive.''  
By contrast, the single reference region that is employed when the conservative method is used is more \juan{widely} separated from Vela\,C, lying at a significantly higher Galactic latitude.
This method may \juan{therefore} under-correct.
Finally, the intermediate diffuse emission subtraction method of \cite{fissel2016} is the mean of the other two methods and was judged to be the most appropriate \juan{approach}.  

Naturally, the use of background subtraction imposes restrictions on the sky areas that may be expected to contain valid data following diffuse emission subtraction.
\cite{fissel2016} define a validity region outside of which the subtraction is shown to be invalid.  
With the exception of North and a very small portion of South-Ridge, all of the \cite{hill2011} sub-regions are included in the validity region.
Unless otherwise specified, we employed the intermediate diffuse emission subtraction \juan{approach}. 
In Appendix~\ref{appendix:refRegions}, we use the aggressive and conservative methods to quantify the extent to which uncertainties associated with diffuse emission subtraction affect our main results.

As noted by \cite{fissel2016}, the point spread function obtained by BLASTPol during our 2012 flight was several times larger than the prediction of our optics model.
Furthermore, the beam was elongated.  
To obtain an approximately round beam, \cite{fissel2016} smoothed their 500\juan{-$\mu$m} data to 2\parcm5 FWHM resolution. 
\cite{gandilo2016} \juan{alternatively} smoothed all three bands to approximately 5\parcm0 resolution in order to compare with \Planck\ results for Vela\,C.
For the purposes of this work, we require similarly shaped and nearly round beams at all three wavelengths, but \juan{we also do not want to sacrifice} resolution.
We were able to achieve these goals by smoothing all three bands to a resolution of 3\parcm0 FHWM.

\subsection{Column density maps}

The column density maps of Vela\,C were derived from the publicly available \Herschel\ SPIRE and PACS data.
SPIRE uses nearly identical filters to BLASTPol, but has higher spatial resolution (FWHM of 17\parcs6, 23\parcs9, and 35\parcs2 for the 250-, 350-, and 500\juan{-$\mu$m} bands, respectively). 
Data taken with the PACS instrument in a band centred at 160\micron\ (FWHM of 13\parcs6) were used to provide additional sensitivity to warm dust.
These \Herschel-based \nh\ maps were generated using {\tt Scanamorphos} \citep{roussel2013} and additional reduction and \juan{data} manipulation \juan{was} performed in the Herschel Interactive Processing Environment ({\tt HIPE} version 11) including the Zero Point Correction function for the SPIRE maps. 
The resulting maps were smoothed to 35\parcs2 resolution by convolving with Gaussian kernels of an appropriate size and then re-gridding to match the \Herschel\ 500\juan{-$\mu$m} map.

We attempted to separate the Galactic foreground and background dust emission from the emission of Vela\,C following the procedure described in section~5 of \cite{fissel2016}.
Modified blackbody spectral energy distribution (SED) fits were made for each map pixel using the methods described in \cite{hill2011} and \juan{with} the dust opacity law presented in \cite{hildebrand1983} \juan{for} a dust spectral index $\beta=2$. 
The result of these fits are column density ($N$) and dust temperature ($T$) maps, both at 35\parcs2 resolution.
It should be noted that above a temperature of approximately 20\,K, the dust emission is expected to peak at wavelengths shorter than 160\micron. For these warmest lines of sight (LOSs) our estimates will have a higher degree of uncertainty. 
Note \juan{also} that we computed maps of the column density of {\it atomic} hydrogen, \nh, while \cite{hill2011} calculated the column density of molecular hydrogen, $N_{\rm H_{2}}$.

% ----------------------------------------------------------------------------------------------------------------------------------------------------------------------

\section{Method}\label{section:analysis}

\subsection{The histogram of relative orientations}

We quantifi\juan{y} the relative orientation between the iso-\nh\ contours and \bperp\ using the histogram of relative orientations \citep[HRO,][]{soler2013}. 
In this technique, the \nh\ structures are characterized by their gradients, which are, by definition, perpendicular to the iso-\nh\ curves. 
The gradient constitutes a vector field that we \juan{can} compare pixel by pixel to the \bperp\ orientation inferred from the polarization maps.

We compute the angle $\phi$ between \bperp\ and the tangent to the \nh\ contours,
\begin{equation}\label{eq:RelativeOrientationAngle}
\phi_{\lambda} = \arctan\left(\,|\,\nabla N_{\rm H} \times \hat{\vec{E}}_{\lambda}\,|, \nabla N_{\rm H}\cdot \hat{\vec{E}}_{\lambda}\,\right).
\end{equation}
For each observation band, characterized by its central wavelength $\lambda$, we assume that \bperp\ is perpendicular to the unit polarization pseudo-vector $\hat{\vec{E}}_{\lambda}$.
The orientation of $\hat{\vec{E}}_{\lambda}$ is defined by the polarization angle $\psi_{\lambda}$ calculated from the observed Stokes parameters using
\begin{equation}
\psi_{\lambda} = \frac{1}{2}\arctan(-U_{\lambda}, Q_{\lambda}).
\end{equation}

In Eq.~\ref{eq:RelativeOrientationAngle}, as implemented, the norm carries a sign when the range used for $\phi_{\lambda}$ is between 0\deg\ and 90\deg.
For the sake of clarity in the representation of the HROs, we cho\juan{o}se the range $0<\phi_{\lambda}<90$\deg, in contrast with the HROs presented in \cite{soler2013} and \cite{planck2015-XXXV}, where $-90<\phi_{\lambda}<90$\deg.
This selection does not imply any loss \juan{of} generality\juan{,} given that the relative orientation is independent of the reference vector, that is, $\phi_{\lambda}$ is equivalent to $-\phi_{\lambda}$.

\subsection{Construction of the histograms}\label{section:hroconstruction}

We construct the HROs using the BLASTPol observations of polarization at 250, 350, and 500\micron\ and the gradient of the total gas column density map, \nh, estimated from the \Herschel\ observations.
The upper panels of Fig.~\ref{fig:HROblastpolherschel} present the maps used in the construction of the HROs. 
For completeness, we present the HROs calculated from the BLASTPol polarization and the gradient of the intensity observed by BLASTPol in the 500\juan{-$\mu$m} band, $I_{500}$, in Appendix~\ref{appendix:HROI}.

The \nh\ HROs have the advantage of providing a more precise estimate of the total gas column density, but the disadvantage of being estimated with a different instrument and at a different angular resolution.
\juan{T}he improvement in angular resolution provides a larger dynamic range for evaluating the HROs at different column densities.

We calculate the gradient of the \nh\ (or $I_{500}$) maps using the Gaussian derivatives method described in \cite{soler2013}.
To guarantee adequate sampling of the derivates in each case, we appl\juan{y} a $5\times5$ derivative kernel computed over a grid with pixel size equal to one third of the beam FWHM in each observation; that is $\Delta l=\Delta b=0$\parcm83 for the BLASTPol $I_{500}$ map (discussed in Appendix~\ref{appendix:HROI}) and $\Delta l=\Delta b=0$\parcm21 for the \Herschel\ \nh\ map.

We compute the relative orientation angle, $\phi_{\lambda}$, introduced in Eq.~\ref{eq:RelativeOrientationAngle}, in all the pixels where BLASTPol polarization observations in each band are available.
We select the polarization observations in terms of their polarized intensity $P_{\lambda}\equiv\sqrt{Q_{\lambda}^2+U_{\lambda}^2}$, such that we only consider $\phi_{\lambda}$ where the polarization signal-to-noise ratios (S/N) $P/\sigma_{P} > 3$. 
These values of the polarization S/N correspond to classical uncertainties in the orientation angle $\sigma_{\psi} <$ 9\pdeg5 and it guarantees that the polarization bias is negligible \citep{serkowski1958,naghizadeh-khouei1993,montier2015}.
 
We compute the HROs in 15 \nh\ bins, each with equal number\juan{s} of $\phi_{\lambda}$ values.
This selection is intended to \juan{examine} the change in $\phi_{\lambda}$ with increasing \nh\ with comparable statistics in each bin.
The lower panels of Fig.~\ref{fig:HROblastpolherschel} present the HROs for the lowest \nh\ bin, an intermediate \nh\ bin, and the highest \nh\ bin. 

\subsection{Relative orientation parameter $\xi$}\label{section:zeta}

The changes in the HROs are quantified using the histogram shape parameter $\xi$, defined as
\begin{equation}\label{eq:zeta}
\xi = \frac{A_{0}-A_{90}}{A_{0}+A_{90}}\, ,
\end{equation}
where $A_{0}$ is the area under the histogram in the range $0\deg< \phi < 22\pdeg5$ and $A_{90}$ is the area under the histogram in the range $67\pdeg5 < \phi < 90\pdeg0$.   
The value of $\xi$ is \juan{nearly} independent of the number of bins selected to represent the histogram if the bin widths are smaller than the integration range.

A histogram peaking at 0\deg, corresponding to \bperp\ mostly aligned with the \nh\ contours, would have $\xi > 0$. 
A histogram peaking at 90\deg, corresponding to \bperp\ mostly perpendicular to the \nh\ contours would have $\xi < 0$. 
A flat histogram, corresponding to no preferred relative orientation between \bperp\ and the \nh\ contours, would have $\xi \approx 0$.

The uncertainty in $\xi$,  $\sigma_{\xi}$, is obtained from
\begin{equation}\label{eq:errzeta}
\sigma^{2}_{\xi} = \frac{4\,(A^{2}_{\rm e}\sigma^{2}_{A_{\rm c}}+A^{2}_{\rm c}\sigma^{2}_{A_{\rm e}})}{(A_{\rm c}+A_{\rm e})^{4}}\, .
\end{equation}
The variances of the areas, $\sigma^{2}_{A_{\rm e}}$ and $\sigma^{2}_{A_{\rm c}}$, characterize the ``jitter'' of the histograms. If the jitter is large, $\sigma_{\xi}$ is large compared to $| \xi |$ and the relative orientation is indeterminate. 
The jitter depends on the number of bins in the histogram, but $\xi$ does not. 

%\subsubsection{Evaluating $\xi$}\label{section:zetaconstruction}

% MOVE THIS PARAGRAPH SOMEWHERE ELSE
We compare the \nh\ maps computed from the \Herschel\ observations with the \bperp\ estimates from the BLASTPol polarization observations.
Given the distance to the cloud, this corresponds to comparing \nh\ structures on scales larger than 0.12\,pc to \bperp\ on scales larger than 0.6\,pc.
This difference in scales implies that we are evaluating the relative orientation of \nh\ structures with respect to a larger-scale component of \bperp.

The \nh\ HROs, shown in the lower panels of Fig.~\ref{fig:HROblastpolherschel}, are flat in the lowest \nh\ \juan{range} and peak at 90\deg\ in the intermediate and highest \nh\ \juan{ranges} across the three BLASTPol wavelength bands.
The HROs corresponding to the highest and intermediate \nh\ \juan{ranges} clearly show fewer counts \juan{for} $0.0<\phi<22.5$ and more \juan{for} $67.5<\phi<90.0$.
%, in contrast with the $I_{500}$ HROs, presented in Appendix~\ref{appendix:HROI}, where the mostly perpendicular relative orientation \juan{is} only \juan{seen for} $67.5<\phi<90.0$.

\begin{figure}[ht!]
\centerline{
\includegraphics[width=0.48\textwidth,angle=0,origin=c]{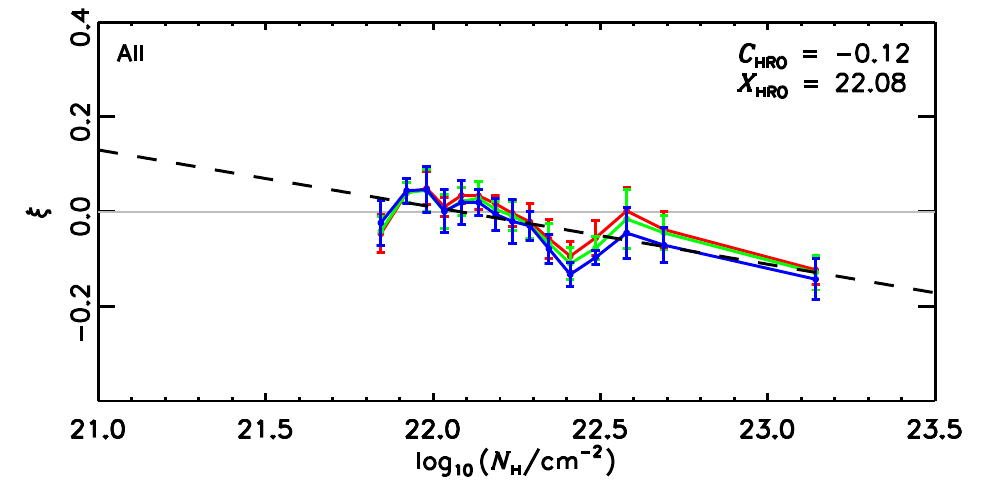}
}
\vspace{-1.0mm}
\caption{Relative orientation parameter $\xi$, defined in Eq.~\ref{eq:zeta}, calculated for the different \nh\ bins towards the South-Nest, South-Ridge, Centre-Nest, and Centre-Ridge \juan{sub-regions of Vela\,C}.
The blue, green, and red curves correspond to the analysis of the BLASTPol observations at 250, 350, and 500\microns, respectively.
The values $\xi > 0$ and $\xi < 0$ correspond to the magnetic field \juan{being} oriented mostly parallel or perpendicular to the iso-\nh\ contours, respectively. 
\juan{The black dashed line and the values of $C_{\textsc{HRO}}$ and $X_{\textsc{HRO}}$ correspond to the linear fit introduced in Eq.~\ref{eq:hrofit}}.
The grey line is $\xi=0$, which corresponds to the case \juan{where} there is no preferred relative orientation.
}
\label{fig:hrozeta}
\end{figure}

Fig.~\ref{fig:hrozeta} \juan{presents} the behaviour of $\xi$ in different \nh\ bins and across the BLASTPol wavelength bands.
The plot shows considerable agreement across the wavelength bands and \juan{shows} that only in the highest \nh\ bin \juan{is} there a clear indication of perpendicular orientation between \nh\ and \bperp\juan{,} while $\xi$ \juan{for} the rest of the \nh\ bins is consistent with no preferred relative orientation.  

As in \cite{planck2015-XXXV}, we characterize the trends in relative orientation \juan{by} assuming \juan{that} the relation between $\xi$ and \lognh\ can be fit roughly by a linear relation
\begin{equation}\label{eq:hrofit}
\xi =  C_{\textsc{HRO}}\,[\log_{10}(\nhd/{\rm cm}^{-2}) - X_{\textsc{HRO}}]\, .
\end{equation}
The measurements \juan{in} each of the BLASTPol wavelength bands are in principle independent determinations of \bperp.
Consequently, we use the estimates of $\xi$ in each wavelength band as independent points in the linear regression that we use to estimate the values of $C_{\textsc{HRO}}$ and $X_{\textsc{HRO}}$.

% -------------------------------------------------------------------------------------------------------------------------------------------------------------------------
\section{Discussion}\label{section:discussion}

We observe that the relative orientation is consistent across the three wavelength bands observed in polarization by BLASTPol. 
This finding \juan{takes} the results of \cite{planck2015-XXXV}, which were based exclusively on 353-GHz (850\juan{-$\mu$m}) polarization observations, \juan{and extends them} not only to a different cloud, but also to the \bperp\ morphology observed in three bands at higher frequencies.
The interpretation of the agreement of the HRO analysis across these three wavelength bands is discussed in Section.~\ref{sec:BandAgreement}.

The results of the \juan{HRO analysis} in Vela\,C suggest that, as in the MCs studied in \cite{planck2015-XXXV}, the magnetic field \juan{plays} a significant role in the assembly of the parcels of gas that become MCs, as \juan{also} suggested by the analysis of simulations of MHD turbulence \citep{soler2013,walch2015,chen2016}.
We discuss this in Section.~\ref{sec:RegionalStudy} \juan{along with} the possible relation between the relative orientation and \juan{the distributions of column density} in different \juan{sub-regions} of the Vela\,C clouds.

% ===================================================================================================
\begin{figure}[ht!]
\centerline{
\includegraphics[width=0.45\textwidth,angle=0,origin=c]{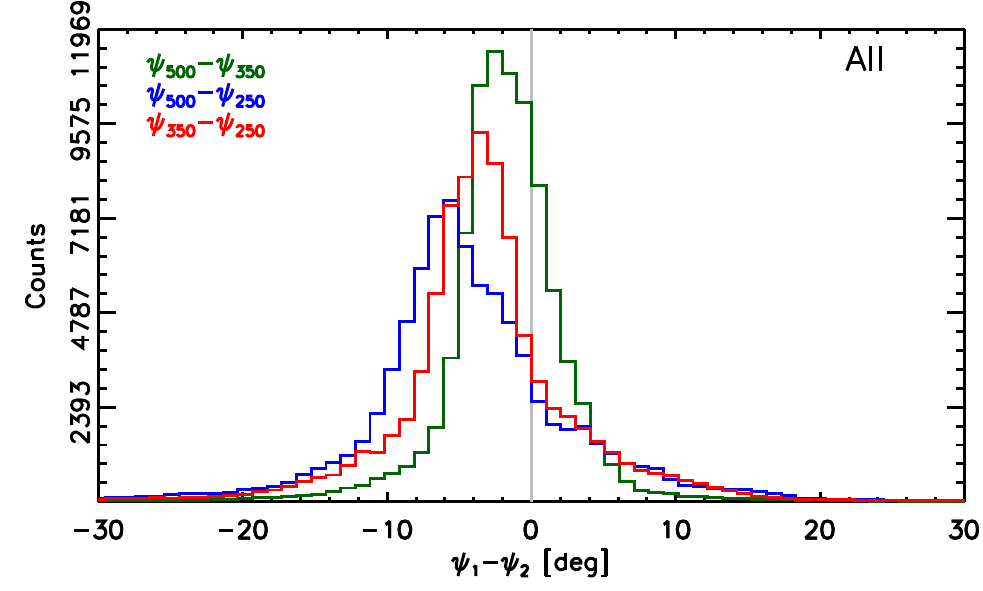}
}
\vspace{-2.0mm}
\centerline{
\includegraphics[width=0.45\textwidth,angle=0,origin=c]{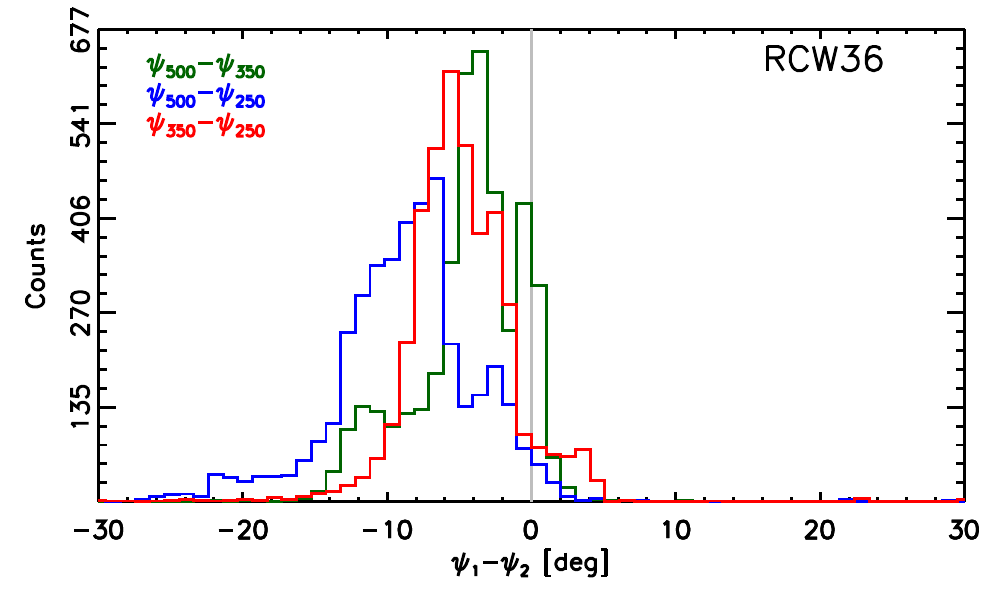}
}
\vspace{-1.0mm}
\caption{Histogram of the polarization angle differences between \juan{the observations at the different BLASTPol wavelength bands towards Vela\,C sub-regions defined in} \cite{hill2011} (top) and in the vicinity of the RCW\,36 region (bottom), that we \juan{have} defined as the portion of the Centre-Ridge sub-region where the dust temperature derived from the \Herschel\ observations is greater than 20\,K, as shown in Fig.~\ref{fig:VelaClumps}.}
\label{fig:AngleDifferences}
\end{figure}

\subsection{Polarization angles \juan{at} different wavelengths}\label{sec:BandAgreement}

Figure~\ref{fig:hrozeta} shows that the relative orientation between the \nh\ structures and \bperp\ inferred from the three BLASTPol wavelength bands is very similar.
This agreement is expe\juan{ct}ed if we consider the small differences between the polarization angles in the different wavelength bands, which we compute directly and present in the top panel of Fig.~\ref{fig:AngleDifferences}.
The average differences between orientation angles in different bands are \juan{all} less than 7\deg, \juan{and} hence have a negligible effect on the shape of the HROs and the behaviour of $\xi$ as function of \nh. 

The most widely accepted mechanism of dust grain alignment, \juan{that of} radiative torques \citep[RATs,][]{lazarian2000}, \juan{explains} the changes in polarization properties across these bands \juan{as arising from} the exposure of different regions to the interstellar radiation field (ISRF) and to internal radiation sources, such as RCW\,36.
According to RATs, dust grains in a cold region deep inside the cloud and far from internal radiation sources will not be as efficiently aligned as the dust grains in regions \juan{where} they can be \juan{spun} up by \juan{a more intense} radiation field.

To investigate whether the polarization properties across the BLASTPol wavelengths depend on the environment in different regions of the clouds, \cite{gandilo2016} evaluated the variations of the polarization fractions, $p_{\lambda}$, in different ranges of \nh\ and temperature, $T$, across the Vela\,C region. 
\cite{gandilo2016} report\juan{ed} that no significant trends of $p_{\lambda}$ were found in different \nh\ ranges, and additionally, no trends over most of the $T$-ranges, except \juan{for} a particular behaviour \juan{for} the highest $T$ data coming from the vicinity of RCW\,36.

\juan{In a similar way}, we \juan{have} evaluated the differences in polarization orientation angles between different BLASTPol wavelength bands in different \nh\ ranges in Vela\,C and summarize the results in Table~\ref{table-angles}, where, for the sake of simplicity, we present only two \nh\ ranges, \juan{namely} \lognh\,$< 22.0$ and \lognh\,$> 22.0$, which are the most relevant for the change in relative orientation between \nh\ \juan{structures} and \bperp.
We observe that the differences in orientation angles between bands are \juan{all} less than 7\deg, and although the differences change \juan{somewhat} with increasing column density, their values do not significantly affect the HROs in the different \nh\ bins.

Towards the bipolar nebula around RCW\,36 \citep{minier2013}, the ionization by the \ion{H}{ii} region and the related increase in dust temperature can potentially introduce differences in the \bperp\ orientation across the BLASTPol wavelength bands.
To test this, we evaluated the difference between the \bperp\ orientation inferred from the observations at different wavelengths in the region around RCW\,36, where the dust temperature, derived from the \Herschel\ observations, is larger than 20\,K.
The results, presented in the bottom panel of Fig.~\ref{fig:AngleDifferences}, indicate that the mean differences in \bperp\ \juan{for} the different BLASTPol wavelength bands are not significantly different \juan{from} those found in the rest of the cloud.
This implies that the magnetic field\juan{,} and the dust that dominates the observed \bperp\ orientations, are approximately the same for \juan{the} three wavelength bands.
This is \juan{not unexpected}, given that the observed \bperp\ is the result of the integration of the magnetic field projection weighted by the dust emission, and 
\juan{most of the dust is in the bulk of}
the cloud, which is most likely unaffected by RCW\,36.

\begin{table*}[ht!]  % table* is a two-column table.  Drop the * for one column.
\begingroup
\newdimen\tblskip \tblskip=5pt
\caption{Differences in polarization angles across BLASTPol wavelength bands in two \lognh\ ranges.}
\label{table-angles}                            % Label goes here.
\nointerlineskip
\vskip -3mm
\footnotesize
\setbox\tablebox=\vbox{
   \newdimen\digitwidth 
   \setbox0=\hbox{\rm 0} 
   \digitwidth=\wd0 
   \catcode`*=\active 
   \def*{\kern\digitwidth}
   \newdimen\signwidth 
   \setbox0=\hbox{+} 
   \signwidth=\wd0 
   \catcode`!=\active 
   \def!{\kern\signwidth}
\halign{\hbox to 1.15in{#\leaderfil}\tabskip 2.2em&
\hfil#&
\hfil#&
\hfil#&
\hfil#&
\hfil#&
\hfil#\tabskip 0pt\cr
\noalign{\doubleline}
\omit\hfil Region$^{a}$\hfil &\multispan3\hfil \lognh\,$< 22.0$ \hfil &\multispan3\hfil \lognh\,$> 22.0$ \hfil \cr
\omit & $\left<\Delta\psi_{500-350}\right>$$^{b}$ & $\left<\Delta\psi_{500-250}\right>$ & $\left<\Delta\psi_{350-250}\right>$ & $\left<\Delta\psi_{500-350}\right>$ & $\left<\Delta\psi_{500-350}\right>$ & $\left<\Delta\psi_{500-350}\right>$ \cr
\noalign{\vskip 4pt\hrule\vskip 6pt}
%----------------------------------------------------------------------------------------------------------------
All regions & \hfil$-0\pdeg7$\hfil & \hfil$-0\pdeg0$\hfil & \hfil\phantom{0}0\pdeg8\hfil & \hfil$-1\pdeg5$\hfil & \hfil$-3\pdeg9$\hfil & \hfil$-2\pdeg5$\hfil \cr
%----------------------------------------------------------------------------------------------------------------
South-Nest & \hfil$\phantom{0}0\pdeg3$\hfil & \hfil\phantom{0}6\pdeg9\hfil & \hfil\phantom{0}6\pdeg6\hfil & \hfil$-2\pdeg1$\hfil & \hfil$-2\pdeg3$\hfil & \hfil$-0\pdeg5$\hfil \cr
%----------------------------------------------------------------------------------------------------------------
South-Ridge & \hfil$-0\pdeg1$\hfil & \hfil$-3\pdeg2$\hfil & \hfil$-3\pdeg0$\hfil & \hfil$-0\pdeg3$\hfil & \hfil$-4\pdeg4$\hfil & \hfil$-4\pdeg0$\hfil \cr
%----------------------------------------------------------------------------------------------------------------
Centre-Nest & \hfil$-0\pdeg2$\hfil & \hfil$-2\pdeg1$\hfil & \hfil$-1\pdeg9$\hfil & \hfil$-0\pdeg7$\hfil & \hfil$-4\pdeg6$\hfil & \hfil$-3\pdeg9$\hfil \cr%----------------------------------------------------------------------------------------------------------------
Centre-Ridge & \hfil$-2\pdeg4$\hfil & \hfil$-5\pdeg2$\hfil & \hfil$-2\pdeg7$\hfil & \hfil$-2\pdeg8$\hfil & \hfil$-5\pdeg0$\hfil & \hfil$-2\pdeg1$\hfil \cr
%----------------------------------------------------------------------------------------------------------------
\noalign{\vskip 3pt\hrule\vskip 4pt}}}
\endPlancktable                    % ends one-column \halign
%\endPlancktablewide                 % ends two-column \halign
\tablenote a As defined in \cite{hill2011} and illustrated in Fig.~\ref{fig:VelaClumps}.\par
\tablenote b $\left<\Delta\psi_{500-350}\right>\equiv\left<\psi_{500}-\psi_{350}\right>$.\par
\endgroup
\end{table*}

% ====================================================================================================
\begin{figure*}[ht!]
\centerline{
\includegraphics[width=0.33\textwidth,angle=0,origin=c]{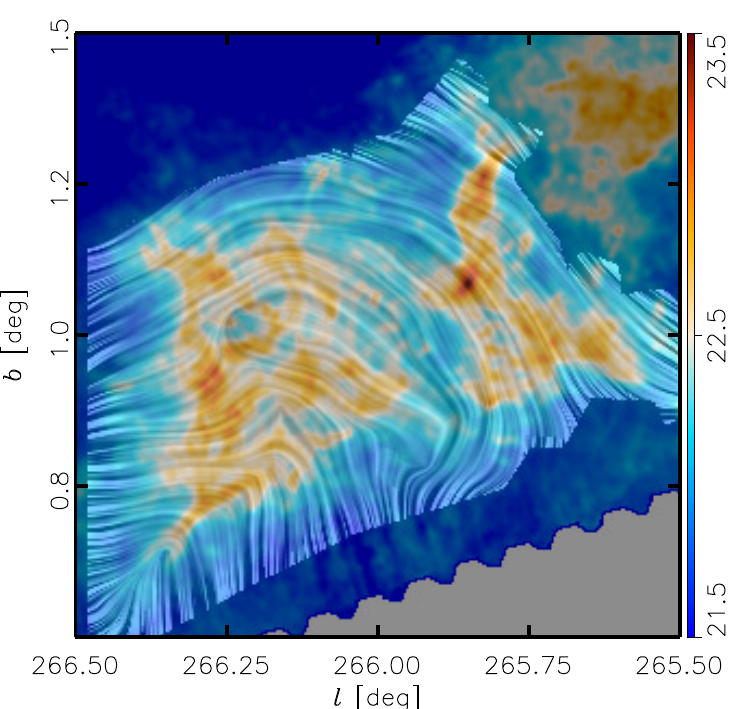}
\includegraphics[width=0.33\textwidth,angle=0,origin=c]{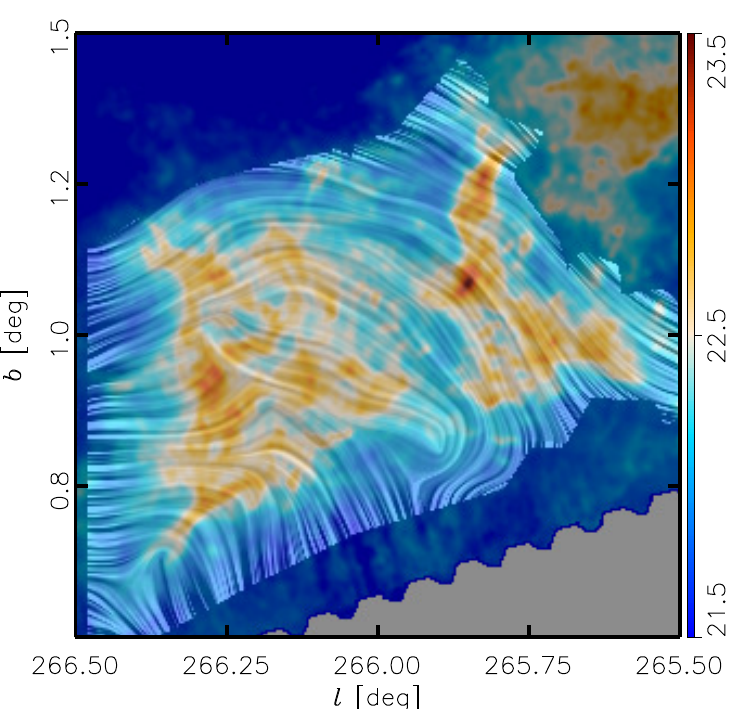}
\includegraphics[width=0.33\textwidth,angle=0,origin=c]{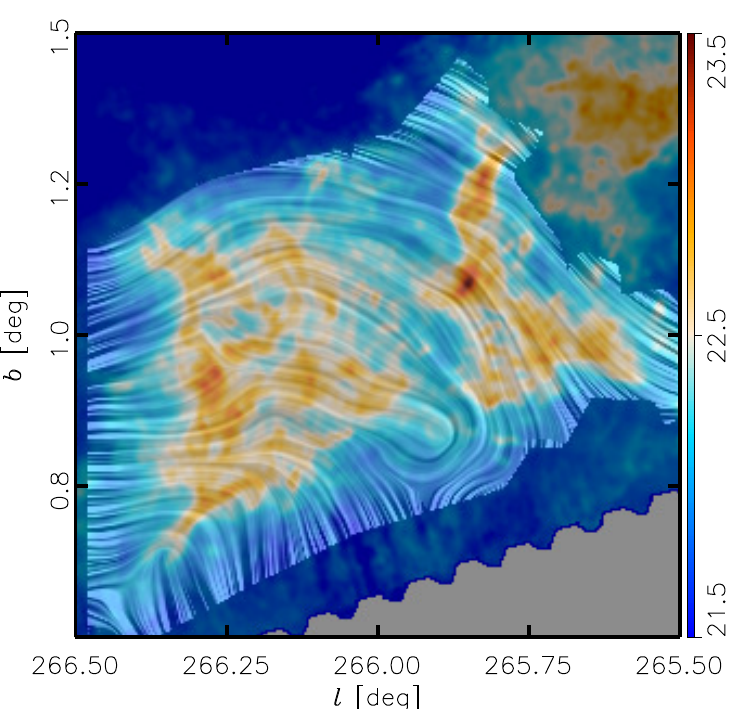}
}
\vspace{-1.0mm}
\centerline{
%\hspace{-0.01\textwidth}\includegraphics[width=0.325\textwidth,angle=0,origin=c]{HROhist250micronSouth-Ridge_int.eps}
%\hspace{0.01\textwidth}\includegraphics[width=0.325\textwidth,angle=0,origin=c]{HROhist350micronSouth-Ridge_int.eps}
%\hspace{0.01\textwidth}\includegraphics[width=0.325\textwidth,angle=0,origin=c]{HROhist500micronSouth-Ridge_int.eps}
\hspace{-0.01\textwidth}\includegraphics[width=0.325\textwidth,angle=0,origin=c]{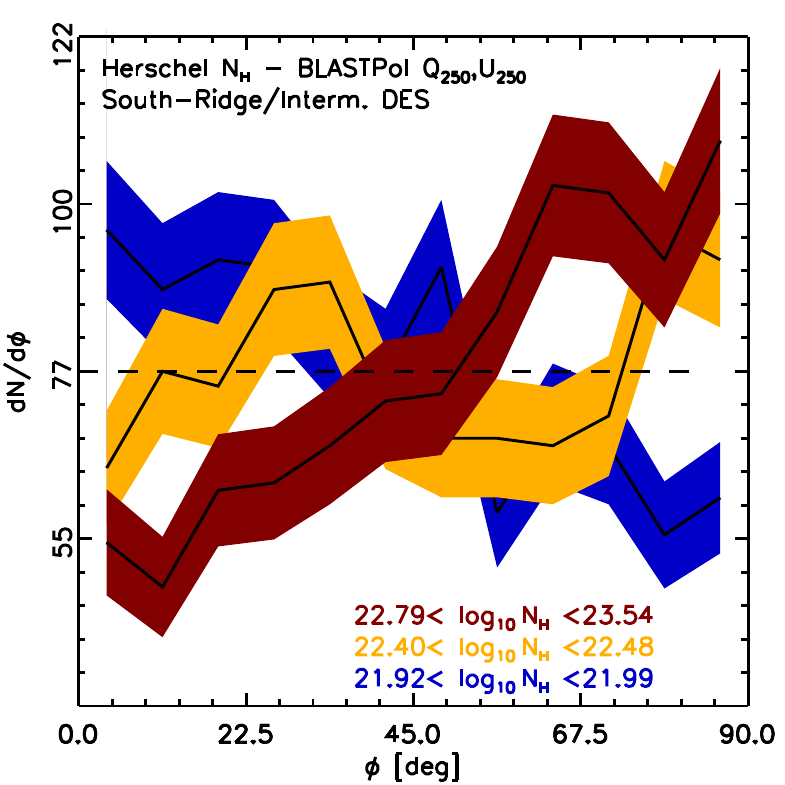}
\hspace{0.01\textwidth}\includegraphics[width=0.325\textwidth,angle=0,origin=c]{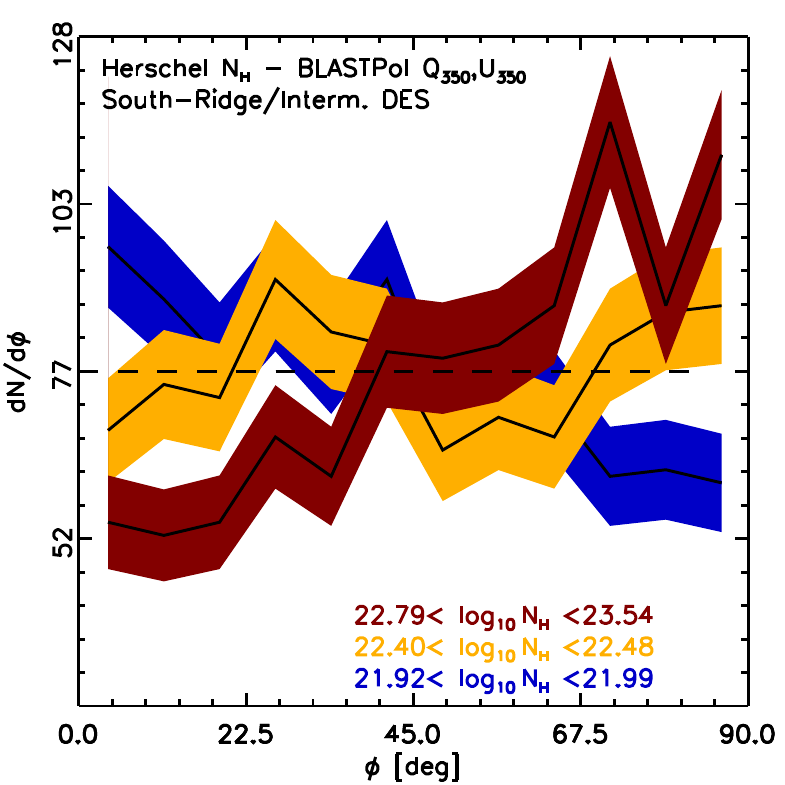}
\hspace{0.01\textwidth}\includegraphics[width=0.325\textwidth,angle=0,origin=c]{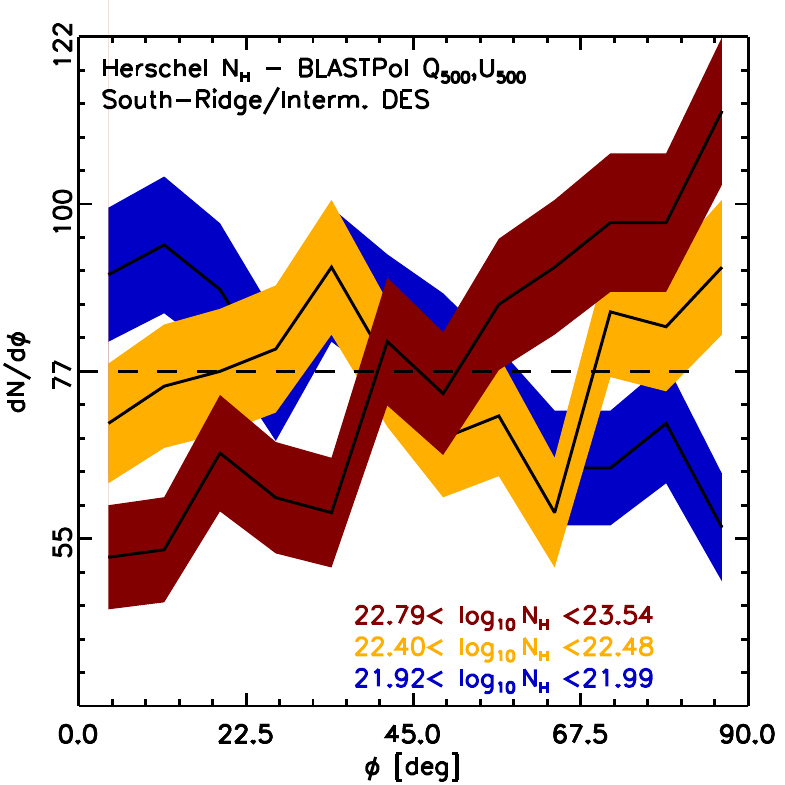}
}
\vspace{-1.0mm}
\centerline{
%\hspace{-0.01\textwidth}\includegraphics[width=0.325\textwidth,angle=0,origin=c]{HROhist250micronSouth-Nest_int.eps}
%\hspace{0.01\textwidth}\includegraphics[width=0.325\textwidth,angle=0,origin=c]{HROhist350micronSouth-Nest_int.eps}
%\hspace{0.01\textwidth}\includegraphics[width=0.325\textwidth,angle=0,origin=c]{HROhist500micronSouth-Nest_int.eps}
\hspace{-0.01\textwidth}\includegraphics[width=0.325\textwidth,angle=0,origin=c]{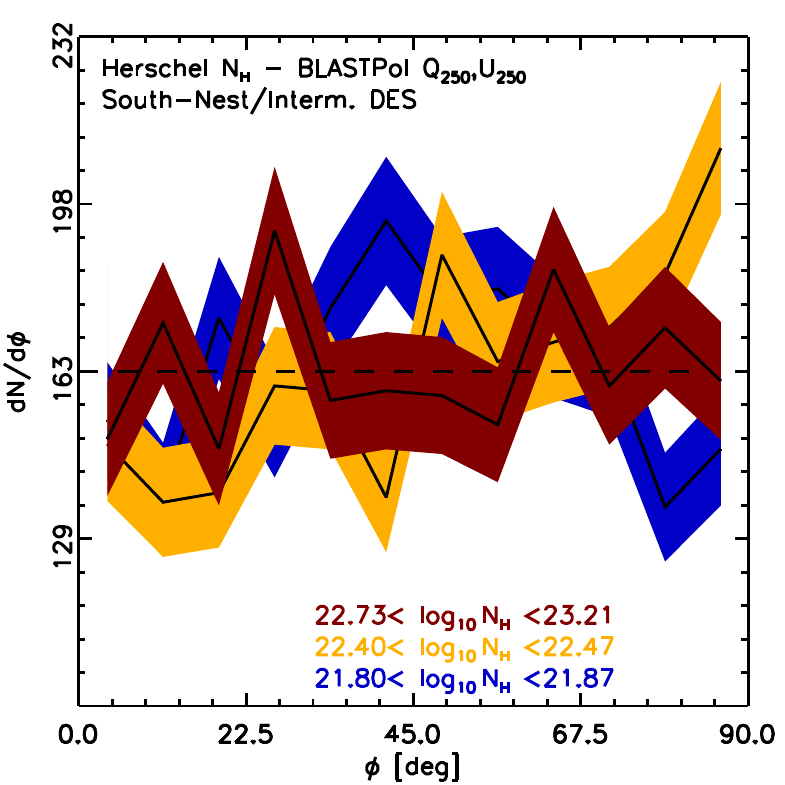}
\hspace{0.01\textwidth}\includegraphics[width=0.325\textwidth,angle=0,origin=c]{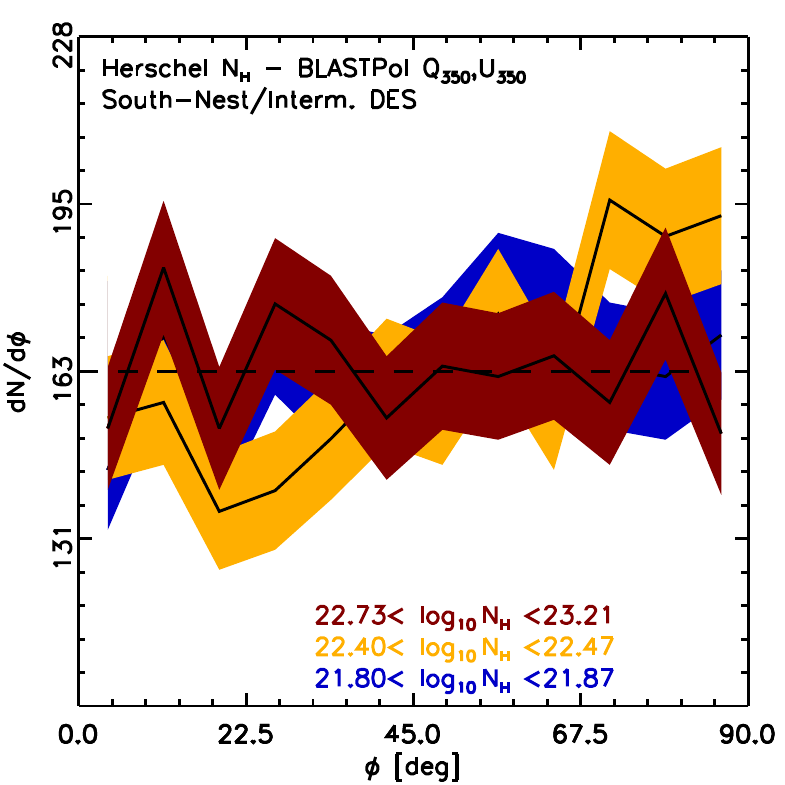}
\hspace{0.01\textwidth}\includegraphics[width=0.325\textwidth,angle=0,origin=c]{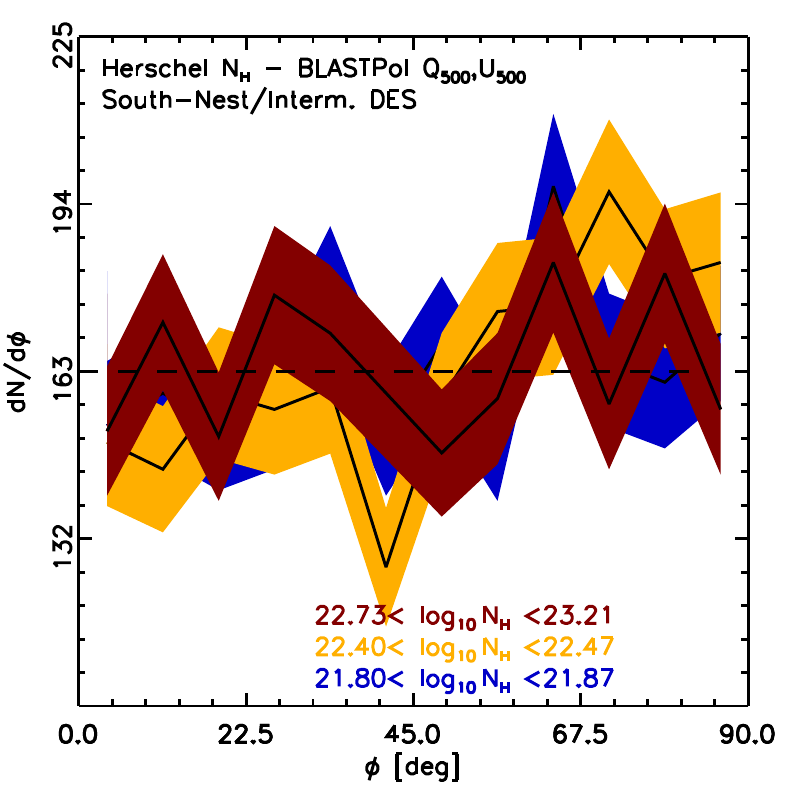}
}
\caption{Same as Fig.~\ref{fig:HROblastpolherschel} for the South-Nest and South-Ridge regions of Vela\,C, as defined in \cite{hill2011}.}
\label{fig:multiHRO_South}
\end{figure*}

\begin{figure*}[ht!]
\centerline{
\includegraphics[width=0.33\textwidth,angle=0,origin=c]{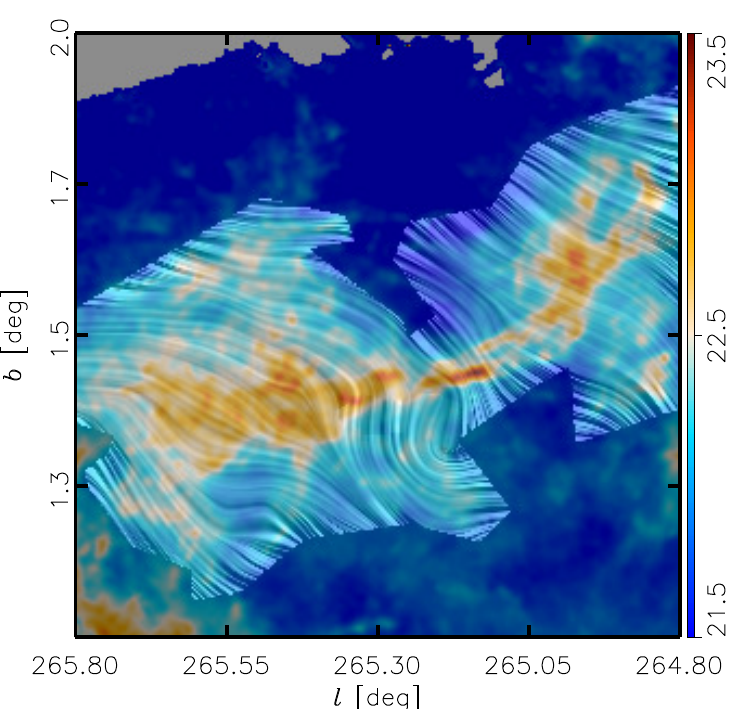}
\includegraphics[width=0.33\textwidth,angle=0,origin=c]{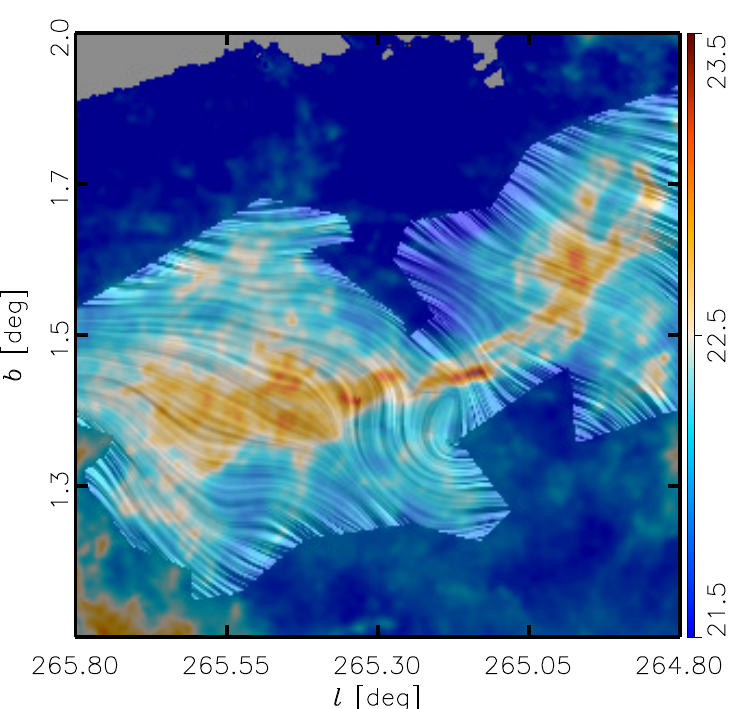}
\includegraphics[width=0.33\textwidth,angle=0,origin=c]{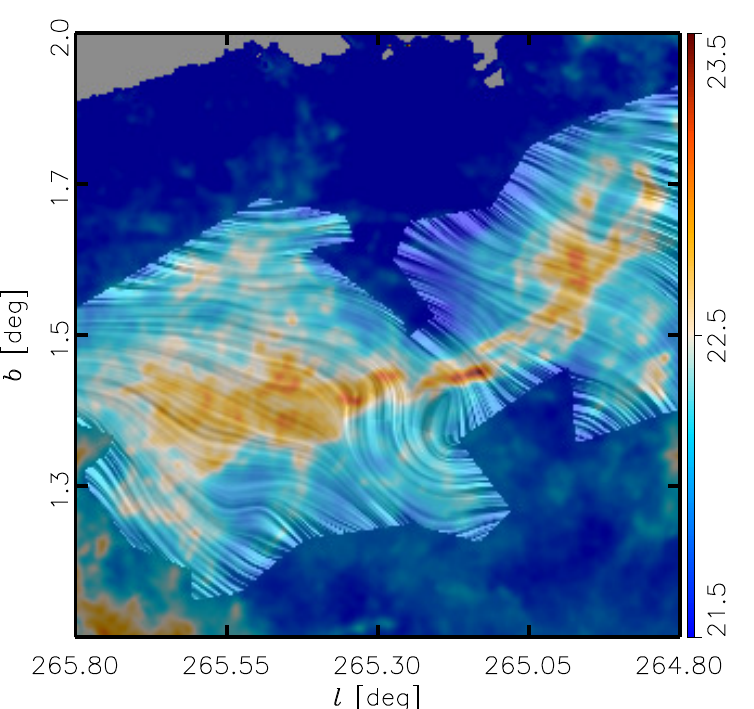}
}
\vspace{-1.0mm}
\centerline{
%\hspace{-0.01\textwidth}\includegraphics[width=0.325\textwidth,angle=0,origin=c]{HROhist250micronCentre-Ridge_int.eps}
%\hspace{0.01\textwidth}\includegraphics[width=0.325\textwidth,angle=0,origin=c]{HROhist350micronCentre-Ridge_int.eps}
%\hspace{0.01\textwidth}\includegraphics[width=0.325\textwidth,angle=0,origin=c]{HROhist500micronCentre-Ridge_int.eps}
\hspace{-0.01\textwidth}\includegraphics[width=0.325\textwidth,angle=0,origin=c]{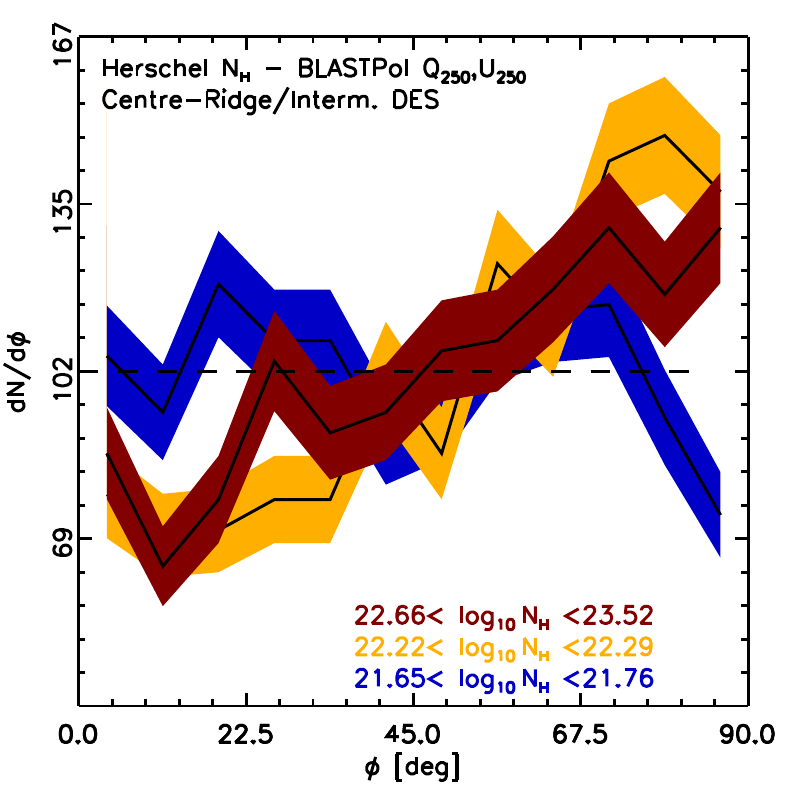}
\hspace{0.01\textwidth}\includegraphics[width=0.325\textwidth,angle=0,origin=c]{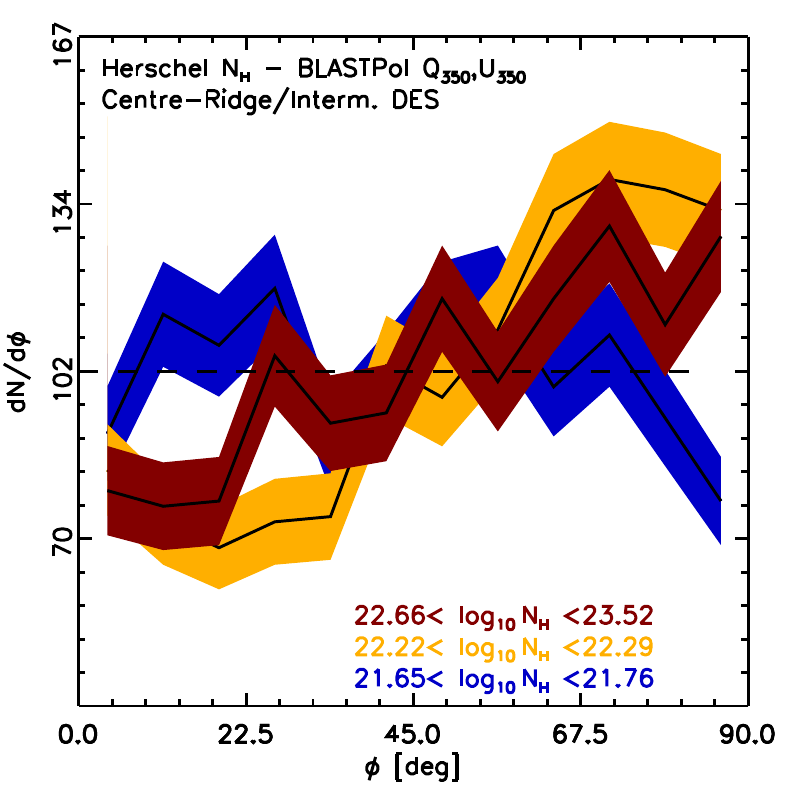}
\hspace{0.01\textwidth}\includegraphics[width=0.325\textwidth,angle=0,origin=c]{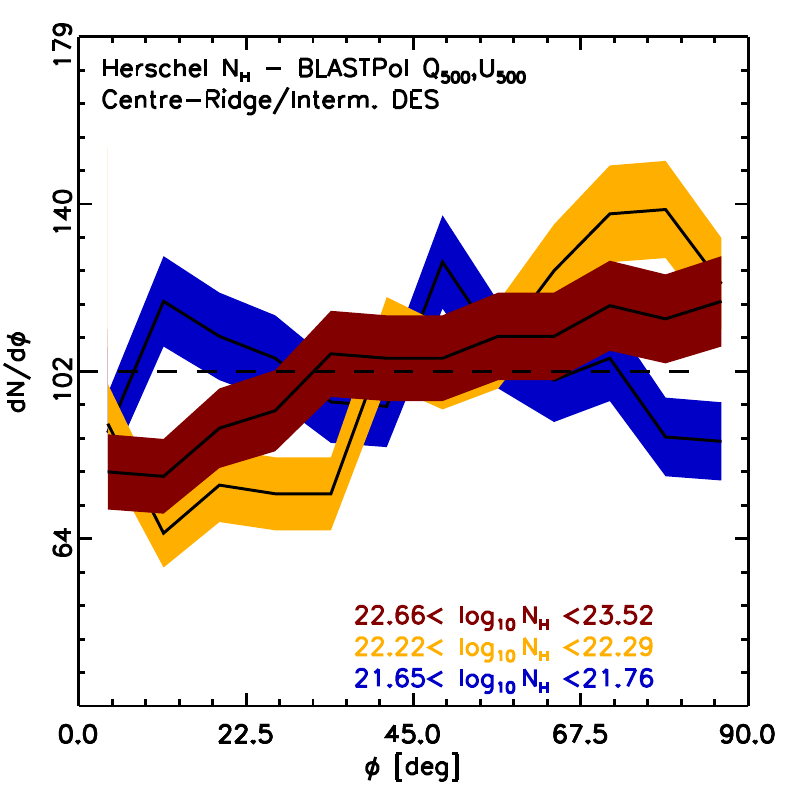}
}
\vspace{-1.0mm}
\centerline{
%\hspace{-0.01\textwidth}\includegraphics[width=0.325\textwidth,angle=0,origin=c]{HROhist250micronCentre-Nest_int.eps}
%\hspace{0.01\textwidth}\includegraphics[width=0.325\textwidth,angle=0,origin=c]{HROhist350micronCentre-Nest_int.eps}
%\hspace{0.01\textwidth}\includegraphics[width=0.325\textwidth,angle=0,origin=c]{HROhist500micronCentre-Nest_int.eps}
\hspace{-0.01\textwidth}\includegraphics[width=0.325\textwidth,angle=0,origin=c]{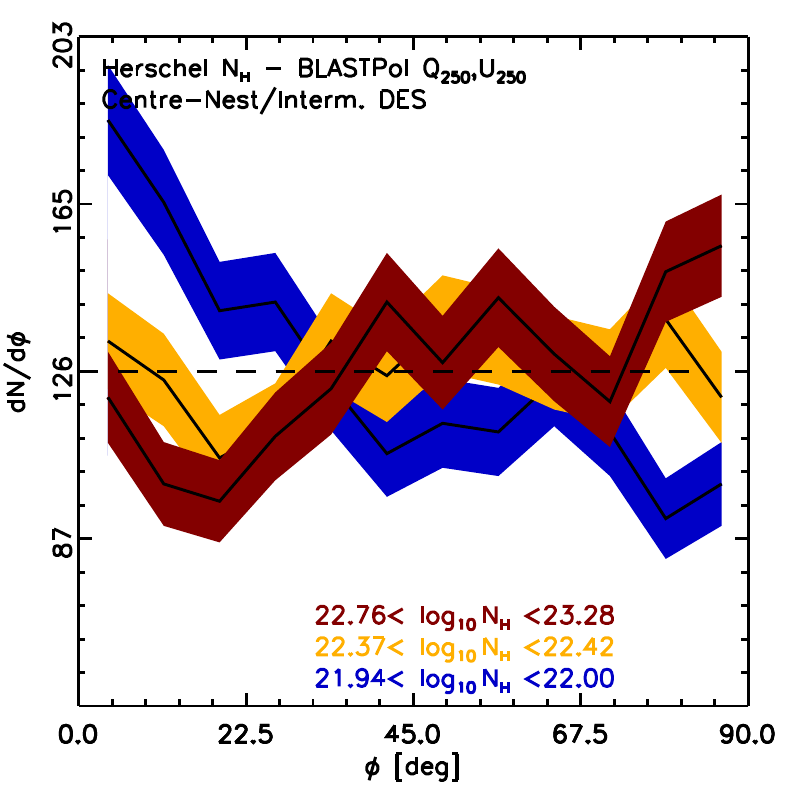}
\hspace{0.01\textwidth}\includegraphics[width=0.325\textwidth,angle=0,origin=c]{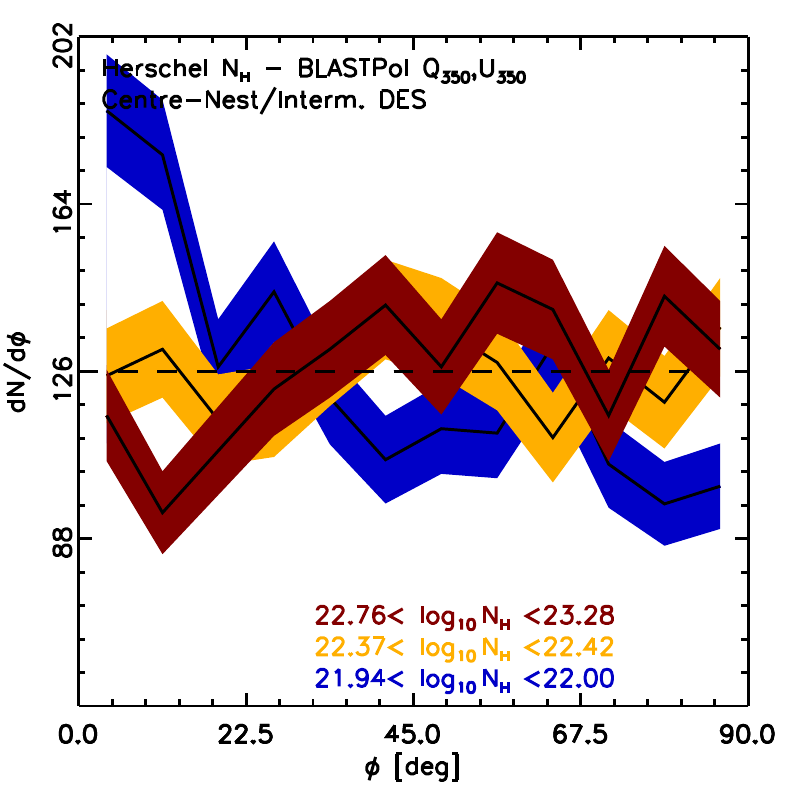}
\hspace{0.01\textwidth}\includegraphics[width=0.325\textwidth,angle=0,origin=c]{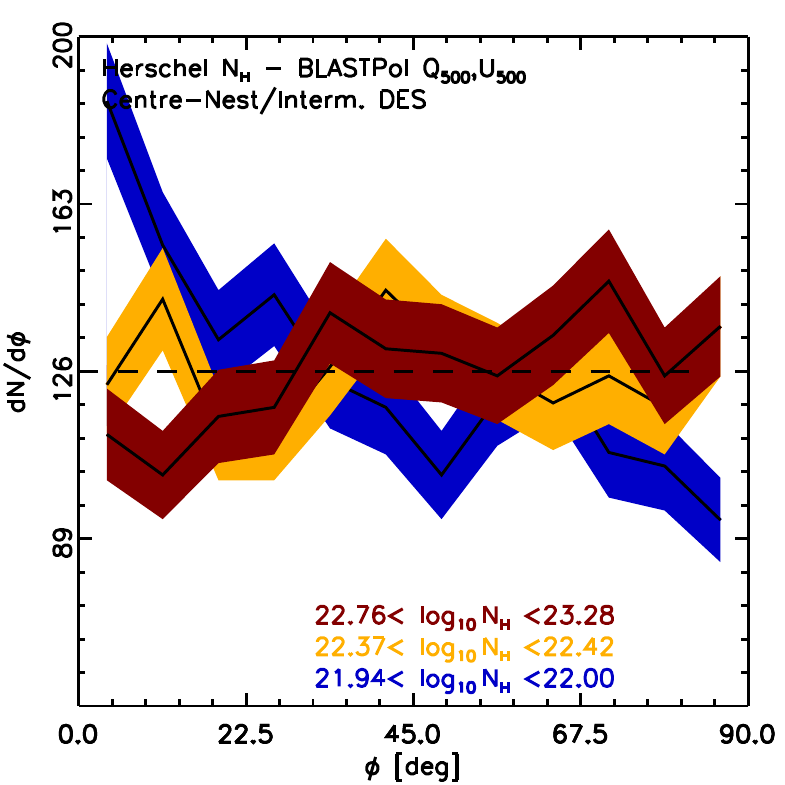}
}
\caption{Same as Fig.~\ref{fig:HROblastpolherschel} for the Centre-Nest and Centre-Ridge regions of Vela\,C as defined in \cite{hill2011}.}
\label{fig:multiHRO_Centre}
\end{figure*}

\subsection{Relative orientations in different portions of the cloud}\label{sec:RegionalStudy}

It is likely that different parts of Vela\,C are at different evolutionary stages; the regions dominated by several filamentary structures with multiple orientations (``nests'') are less evolved, and regions dominated by one clear filamentary structure (``ridges'') are in a more advanced state of evolution.
To test this hypothesis\juan{,} in terms of the relative orientation between the \nh\ structures and \bperp, we computed the HROs separately in four sub-regions of Vela\,C, namely South-Ridge, Centre-Ridge, South-Nest, and Centre-Nest, as identified in \cite{hill2011} and illustrated in Fig.~\ref{fig:VelaClumps}.

The maps of the South and Centre portions of Vela\,C are presented in the top panels of Fig.~\ref{fig:multiHRO_South} and Fig.~\ref{fig:multiHRO_Centre}, respectively.
The middle and bottom panels of Fig.~\ref{fig:multiHRO_South} and Fig.~\ref{fig:multiHRO_Centre} show the HROs that correspond to the ridge and nest sub-regions of the South and Centre portions of Vela\,C.
\juan{The comparison of these HROs indicates that there is a clear difference in relative orientations at the lowest \nh\ and the highest \nh\ bins in all of the sub-regions except for the South-Nest.

In the Centre-Ridge and the South-Ridge}, the HRO corresponding to \juan{the} lowest \nh\ bin is flat or slightly peak\juan{s} around 0\deg, in contrast with the HROs in the intermediate and highest \nh\ bins, which peak at 90\deg.
In the Centre-Nest region, the HROs at the highest and intermediate \nh\ bins are relatively flat, but the HRO in the lowest \nh\ bin shows more counts in the $0<\phi_{\lambda}<22.5$\deg\ range than in $67.5<\phi_{\lambda}<90$\deg, suggesting alignment between the lowest \nh\ contours and \bperp\ in this sub-region.
In the South-Nest region, the HROs present a large amount of jitter\juan{, which} makes it difficult to identify any preferential relative orientation, but the histogram in the intermediate \nh\ bin clearly shows fewer counts in the $0<\phi_{\lambda}<22.5$\deg\ range than in $67.5<\phi_{\lambda}<90$\deg, suggesting that the \nh\ contours are mostly perpendicular to \bperp\ at the highest \nh\ in this sub-region.
\juan{However,} a more effective evaluation of the change in relative orientation between the iso-\nh\ contours and \bperp\ is made in terms of the values of $\xi$, which we show in Fig.~\ref{fig:hrozetaNRmulti}.

\subsubsection{Nests vs. ridges}\label{sec:nestvsridges}

It is clear from the values of $\xi$ in Fig.~\ref{fig:hrozetaNRmulti} that the change in relative orientation, from mostly parallel \juan{or having no preferred orientation} to mostly perpendicular with increasing \lognh, is more evident in the ridge regions.
There is some indication of \juan{a similar} change in relative orientation in the nest regions, but there the values of $\xi$ are closer to zero  \juan{at the high-column-density end}.

In terms of the linear fit used to characterize the behaviour of $\xi$ as function of \lognh, \juan{introduced} in Eq.~\ref{eq:hrofit}, all of the sub-regions present negative values of the slope, $C_{\textsc{HRO}}$, suggesting the change from $\xi>0$ (or \nh\ contours and \bperp\ mostly parallel) to $\xi<0$ (or \nh\ contours and \bperp\ mostly perpendicular) with increasing \lognh, although this transition is only completely \juan{clear in} the observations towards the South-Ridge region, which also presents \juan{the} most negative value of $C_{\textsc{HRO}}$.
The values of $C_{\textsc{HRO}}$, also summarized in Table~\ref{table-zeta}, are comparable to those \juan{reported} in \cite{planck2015-XXXV} for 10 nearby MCs, where the steepest slope was found towards IC 5146 and the Aquila rift, with $C_{\textsc{HRO}}=-0.68$ and $-0.60$ respectively, and the shallowest towards the Corona Australis, \juan{with} $C_{\textsc{HRO}}=-0.11$ as shown in table~2 of that reference.

The values of $X_{\textsc{HRO}}$, which roughly correspond to the \lognh\ values where $\xi$ changes from positive to negative, or where the relative orientation between the \nh\ contours and \bperp\ changes from mostly parallel to mostly perpendicular, are also comparable with those in \cite{planck2015-XXXV}.
The values of $X_{\textsc{HRO}}=22.71$, $22.39$, and $22.18$ seen towards the Centre-Nest, the South-Ridge, and the South-Nest, respectively, are comparable to those found in Ophiuchus, $X_{\textsc{HRO}}=22.70$, and the Aquila rift, $X_{\textsc{HRO}}=22.23$.
A particular case seems to be the Centre-Ridge, where $\xi$ is clearly less tha\juan{n} zero for most \lognh\ values and oscillates around $-0.2$ for \lognh\,$<21.7$, making the values of $X_{\textsc{HRO}}$ and $C_{\textsc{HRO}}$ not very informative. 
\juan{However, the mostly negative values of $\xi$ towards this regions are significant, given} that the Centre-Ridge includes RCW\,36 and, as commented by \cite{hill2011}, it contains most of the massive cores in Vela\,C.

\juan{When interpreting those results, i}t is possible that the different trends in the relative orientation between the \nh\ contours and \bperp\ presented in Fig.~\ref{fig:hrozetaNRmulti} correspond to different projection effects in each one of the sub-regions, that is, the mean magnetic field has a different orientation with respect to the LOS\juan{,} directly affecting the behaviour of $\xi$ as a function of \lognh.
From geometrical arguments, explained in detail in Appendix C of \cite{planck2015-XXXV}, it is unlikely that the relative orientation in the South-Ridge and Centre-Ridge is too far from the three-dimensional configuration of the magnetic field and the density structures.
\juan{But it is possible} that the values of $\xi$ around zero in the South-Nest and Centre-Nest are produced by the mean magnetic field \juan{being} oriented close to the LOS, thus making it difficult to identify a preferential relative orientation between the projected density structure and \bperp.
Alternatively, it is possible that the South-Nest and Centre-Nest are more turbulent, as suggested by the dispersion in \bperp\ orientations inferred from the near-infrared polarimetry observations presented in \cite{kusune2016}. 
However, as described in the same reference, this hypothesis is not clearly justified by the width of $^{13}$CO emission lines, which are not correlated with the \bperp\ dispersions in  the sub-regions. 

\juan{Detailed} comparison between the \bperp\ morphology and the velocity information inferred from different molecular tracer\juan{s} is beyond the scope of this work, but will be considered in a forthcoming publication (Fissel et al. 2017).
For the moment, we \juan{simply} compare the implications of the different relative orientations between \nh\ contours and \bperp\ in the sub-regions of Vela\,C, and \juan{consider this in} relation to other cloud characteristics, such as the \juan{column density probability distribution functions}.

\begin{figure}[ht!]
\centerline{
\includegraphics[width=0.49\textwidth,angle=0,origin=c]{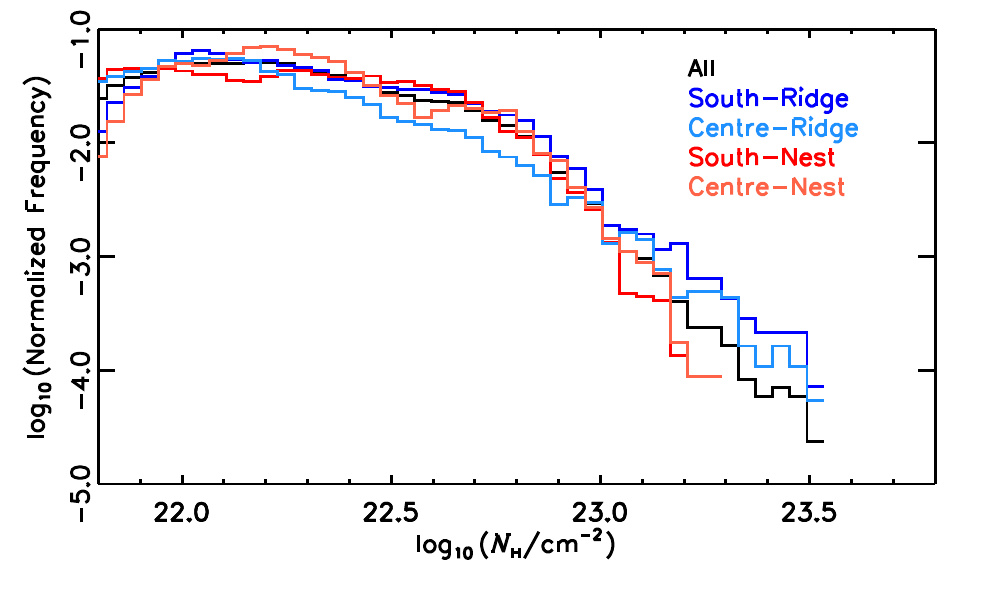}
}
\vspace{-3.0mm}
\caption{\juan{Comparison of the normalized \nh\ probability distribution functions (PDFs), inferred from the \Herschel\ observations towards the four sub-regions of Vela\,C presented in Fig.~\ref{fig:VelaClumps}.
The y-axis is presented in logarithmic scale for the sake of comparison with figure~5 of \cite{hill2011}.}
}
\label{fig:VelaCNHPDF}
\end{figure}

\subsubsection{Column density probability distribution functions and relative orientations}

\cite{hill2011} describe a significant difference between the ridge and nest sub-regions of Vela\,C: while all the clouds have comparable masses, the highest values of the column density probability distribution functions (PDFs) are found in the ridge sub-regions, as illustrated in Fig.~\ref{fig:VelaCNHPDF}. 
Moreover, \juan{\cite{hill2011} also indicate that} most of the cores with $M>8$\,M$_\odot$ are found in the Centre-Ridge region, while the Centre-Nest has two, the South-Ridge has one, and the South-Nest has none.
\juan{This observation is confirmed by the distribution of dense cores in Vela\,C presented in \cite{giannini2012}, as illustrated in Fig.~\ref{fig:VelaClumps}}.
\cite{hill2011} interpret this as the effect of ``constructive large-scale flows'' and strong compression of material in the Centre-Ridge.
The results of the HRO analysis indicate that such flows \juan{are} parallel to the mean magnetic field direction, resulting in the high column density structure that is perpendicular to the magnetic field.

\juan{To test this hypothesis in a different region}, we evaluated \juan{the column density PDFs and} the distribution of massive cores and the relative orientation between \nh\ and \bperp\ in the Serpens South sub-region of the Aquila complex, \juan{as described in detail in Appendix~\ref{appendix:AquilaRegion}}.
Serpens South is slightly less massive that Vela\,C, but it is less affected by feedback than Orion, which is the closest mass equivalent in the group of clouds studied in \cite{planck2015-XXXV}.
The observations presented in \cite{hill2012} indicate that the filamentary structure in the Vela\,C Centre-Ridge and the Serpens South filament are quite similar in terms of their column density profiles, and mass per unit length.
Additionally, they are similar in that both contain an \ion{H}{ii} region, RCW\,36 in the case of Vela\,C and W40 in the case of Serpens South.

The HRO analysis of Serpens South indicates that as in Vela\,C, \juan{the maximum column densities, the flattest slopes of the column density PDF tail, and} the largest number of  \juan{dense cores are located in the portion of the cloud with the sharpest transition in the relative orientation between gas column density structures and the magnetic field.}
The transition is from \bperp\ being mostly parallel or having no preferred alignment with respect to the \nh\ structures to mostly perpendicular; see Sect.~\ref{sec:nestvsridges} and Appendix~\ref{appendix:AquilaRegion}.
This observational fact, which \juan{alone does not imply causality}, \juan{is significant if we consider the current observations of the assembly of density structures in molecular clouds \citep[][and the references therein]{andre2014}.}

\begin{table*}[ht!]  % table* is a two-column table.  Drop the * for one column.
\begingroup
\newdimen\tblskip \tblskip=5pt
\caption{Parameters of the relative orientation between \nh\ and \bperp\ towards the Vela\,C sub-regions}
\label{table-zeta}                            % Label goes here.
\nointerlineskip
\vskip -3mm
\footnotesize
\setbox\tablebox=\vbox{
   \newdimen\digitwidth 
   \setbox0=\hbox{\rm 0} 
   \digitwidth=\wd0 
   \catcode`*=\active 
   \def*{\kern\digitwidth}
   \newdimen\signwidth 
   \setbox0=\hbox{+} 
   \signwidth=\wd0 
   \catcode`!=\active 
   \def!{\kern\signwidth}
\halign{\hbox to 1.15in{#\leaderfil}\tabskip 2.2em&
\hfil#&
\hfil#&
\hfil#&
\hfil#\tabskip 0pt\cr
\noalign{\doubleline}
\omit\hfil Region\hfil & \hfil$C_{\textsc{HRO}}$$^{a}$\hfil &  \hfil$X_{\textsc{HRO}}$$^{a}$\hfil & \hfil Max($\log$(\nh/cm$^{-2}$))$^{b}$\hfil & \hfil Mean($\log$(\nh/cm$^{-2}$))$^{b}$\hfil \cr
%\omit & & & \hfil[10$^{22}$\,cm$^{-2}$]\hfil & \hfil[10$^{22}$\,cm$^{-2}$]\hfil \cr
\noalign{\vskip 4pt\hrule\vskip 6pt}
%----------------------------------------------------------------------------------------------------------------
South-Nest &  \hfil$-0.11$\hfil & \hfil22.23\hfil & \hfil22.9\hfil & \hfil22.1\hfil \cr
%----------------------------------------------------------------------------------------------------------------
South-Ridge & \hfil$-0.38$\hfil & \hfil22.40\hfil & \hfil23.3\hfil & \hfil22.2\hfil \cr
%----------------------------------------------------------------------------------------------------------------
Centre-Nest &  \hfil$-0.17$\hfil & \hfil22.62\hfil & \hfil23.0\hfil & \hfil22.2\hfil \cr
%----------------------------------------------------------------------------------------------------------------
Centre-Ridge & \hfil$-0.12$\hfil & \hfil20.69\hfil & \hfil23.2\hfil & \hfil22.1\hfil \cr
%----------------------------------------------------------------------------------------------------------------
\noalign{\vskip 3pt\hrule\vskip 4pt}}}
\endPlancktable
\tablenote a Fit of $\xi$ vs.\ \lognh. See Eq.~\ref{eq:hrofit} and Fig.~\ref{fig:hrozetaNRmulti}.\par
\tablenote b From \cite{hill2011}.\par
\endgroup
\end{table*}  
\begin{figure*}[ht!]
\centerline{
\includegraphics[width=0.5\textwidth,angle=0,origin=c]{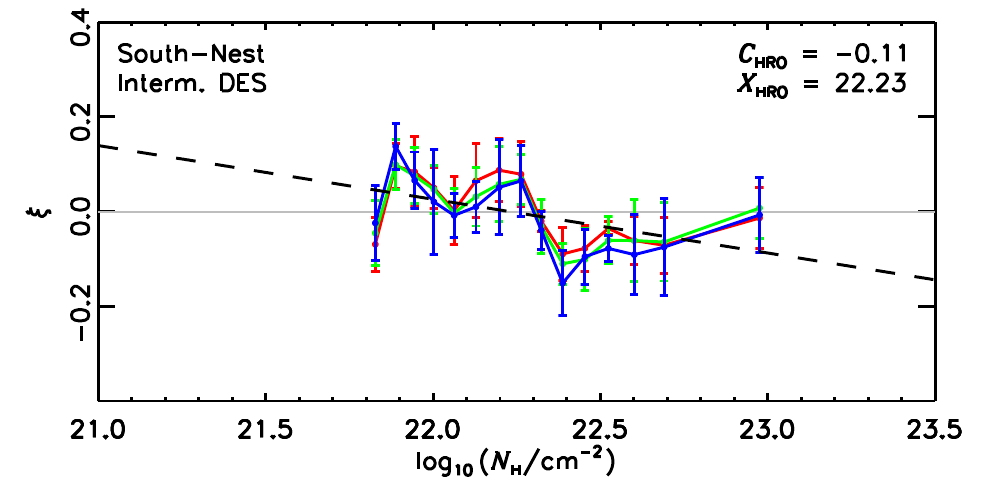}
\includegraphics[width=0.5\textwidth,angle=0,origin=c]{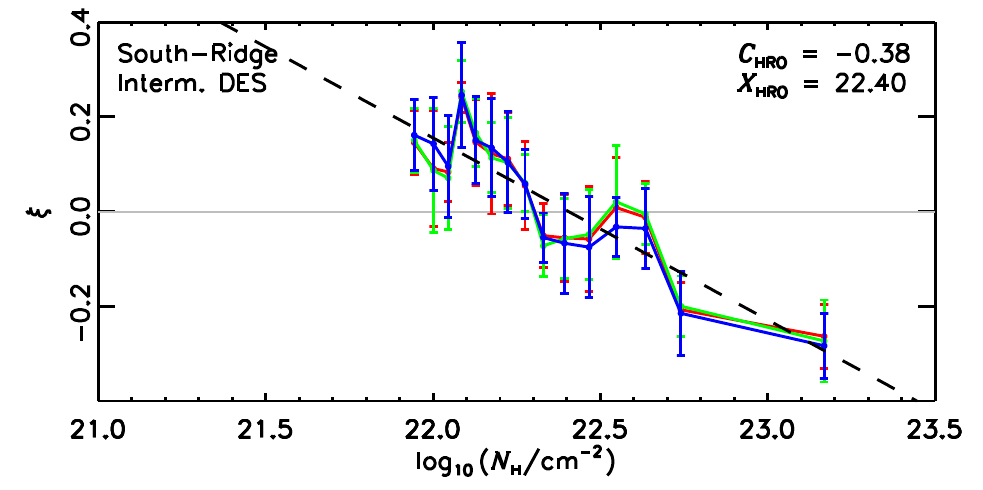}
}
\vspace{-2.0mm}
\centerline{
\includegraphics[width=0.5\textwidth,angle=0,origin=c]{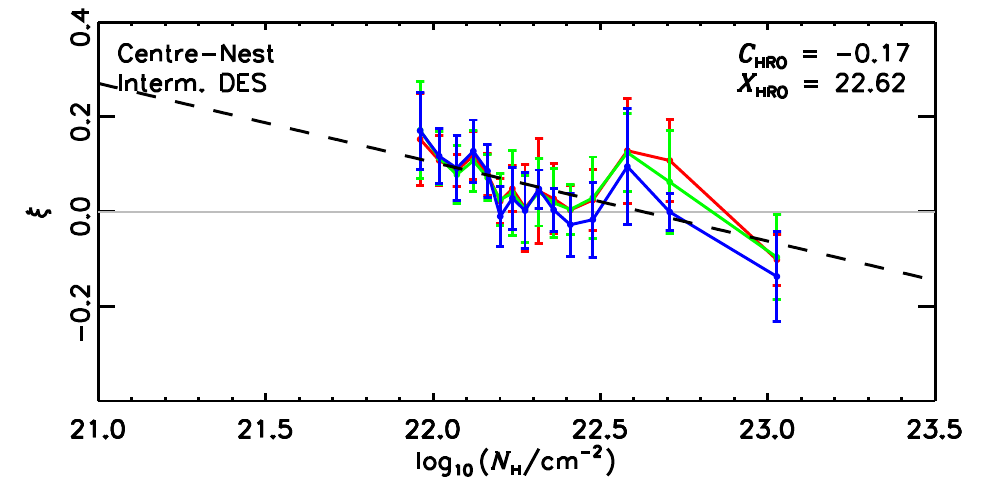}
\includegraphics[width=0.5\textwidth,angle=0,origin=c]{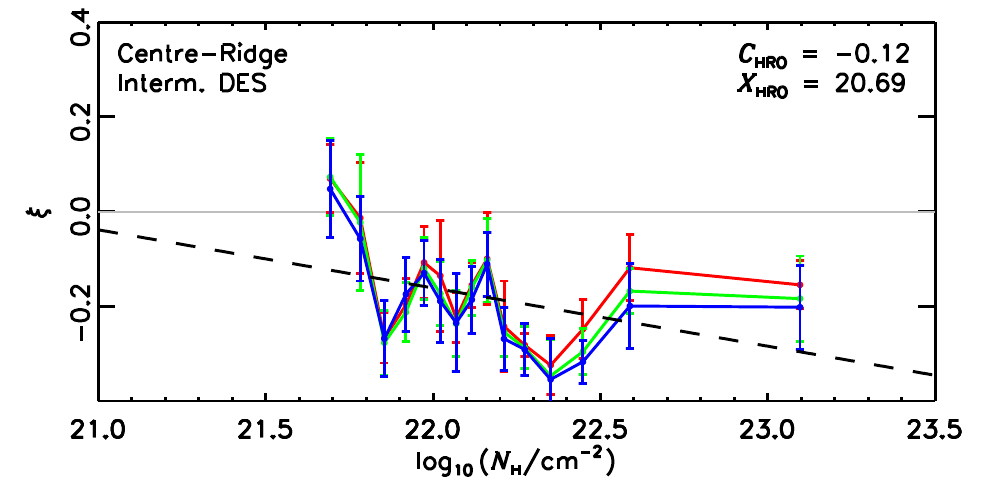}
}
\vspace{-1.0mm}
\caption{Relative orientation parameter $\xi$, defined in Eq.~\ref{eq:zeta}, calculated for the different \nh\ bins towards the sub-regions of Vela\,C as defined in \cite{hill2011} and illustrated in Fig.~\ref{fig:VelaClumps}. 
The values $\xi > 0$ and $\xi < 0$ correspond to the magnetic field \juan{being} oriented mostly parallel \juan{or} perpendicular to the iso-\nh\ contours, respectively. 
\juan{The black dashed line and the values of $C_{\textsc{HRO}}$ and $X_{\textsc{HRO}}$ correspond to the linear fit introduced in Eq.~\ref{eq:hrofit}}.
The grey line is $\xi=0$, which corresponds to the case \juan{where} there is no preferred relative orientation.
}
\label{fig:hrozetaNRmulti}
\end{figure*}

\subsection{The role of the magnetic field in molecular cloud formation}

Each of the sub-regions of Vela\,C contains roughly the same mass \citep{hill2011}.
Assuming a constant star-formation efficiency, each sub-region should \juan{therefore} form approximately the same number of massive stars, \juan{which is not what is observed in terms of the column density PDF and high-mass core distribution}.
\juan{Then, the difference between the sub-regions may be in their efficiency for gathering material into the dense regions where star formation is ongoing.}

\juan{\cite{kirk2013} and \cite{palmeirim2013} presented observational evidence of the potential feeding of material into hubs or ridges, in high- and low- mass star forming regions respectively, where the gathering flows seem to follow the magnetic field direction.
The finding of ridge-like structures towards the Centre-Ridge and the South-Ridge, where the \nh\ structures are mostly perpendicular to \bperp, suggest that these sub-regions in Vela\,C were formed through a similar mechanism}.

\cite{planck2015-XXXV} \juan{argue} that the transition between mostly parallel and mostly perpendicular is related to the balance between the kinetic, gravitational, and magnetic energies and the flows of matter, which are restricted by the Lorentz force to follow the magnetic fields.
A frozen-in and strong interstellar magnetic field would naturally cause a self-gravitating, static cloud to become oblate, with its major axis perpendicular to the field lines, because gravitational collapse would be restricted to occurring along field lines \citep{mestel1956,mestel1984,mouschovias1976I}.

In a dynamic picture of MCs, the matter-gathering flows, such as the ones driven by expanding bubbles \citep{inutsuka2015}, are more successful at gathering material and eventually forming MCs if they are close to parallel to the magnetic field \citep{hennebelle2000,ntormousi2017}. 
Less dense structures, \juan{which} are not self-gravitating, would be stretched along the magnetic field lines by velocity shear, thereby producing aligned density structures, as discussed in \cite{hennebelle2013a}.
\cite{walch2015} report a similar effect \juan{for} the magnetic field in a 500\,pc simulation of a \juan{G}alactic disc including self-gravity, magnetic fields, heating and radiative cooling, and supernova feedback.
In the evolution of their simulation setup, they observe that the magnetic field is delaying or favouring the collapse in certain regions by changing the amount of dense and cold gas formed.
The additional magnetic pressure is significant in dense gas and thus \juan{slows} the formation of dense and cold, molecular gas.

\juan{Given the aforementioned observational and theoretical considerations, it is plausible to consider that the Centre-Ridge is a more evolved region, where the structures produced by the flows along \bperp, which would be mostly parallel to \bperp, have already collapsed into the central object. 
This structure is gravitationally bound and collapsing into dense sub-structures, including high-mass stars, such as those found in RCW\,36.
This would imply that the South-Ridge is in an early state of accretion, where the column density structures that are mostly oriented along \bperp, at \lognh\,$<22.6$, are the product of the inflows feeding the highest column density structures.
In this scenario, the Center-Nest and South-Nest sub-regions would correspond to regions that are less efficient in gathering material due to interaction between matter-gathering flows and the magnetic field.}

% -------------------------------------------------------------------------------------------------------------------------------------------------------------------------
\section{Conclusions}\label{section:conclusions}

We \juan{have} evaluated the relative orientations between the column density structure, \nh\juan{,} inferred from the \Herschel\ satellite observations, and the magnetic field projected on the plane of sky, \bperp\juan{,} inferred from observations of polarization at 250, 350, 500\micron\ by \BLASTPol, towards the Vela\,C molecular complex.
We found that the relative orientation between the iso-\nh\ contours and \bperp, changes progressively with increasing \nh, from preferentially parallel or having no preferred orientation to preferentially perpendicular, in agreement with the behaviour reported by \cite{planck2015-XXXV} in ten nearby MCs.

We found close agreement between the values of the \bperp\ orientation inferred from observations in the three BLASTPol wavelength bands.
Consistently, we found very similar trends in relative orientation between the iso-\nh\ contours and \bperp\ in the three BLASTPol wavelength bands.
This agreement, together with the flat polarization SED across the bands reported in \cite{gandilo2016} suggest that the observed \bperp\ orientation  
\juan{is dominated by regions in the bulk of the gas associated \juan{with} the MC, which contain most of the dust and are exposed to a relatively homogeneous radiative environment that, at the considered scales, is less likely to produce significant changes in the dust grain alignment}.

\juan{We studied the relative orientations between the iso-\nh\ contours and \bperp\ towards different sub-regions of Vela\,C.
We found different behaviours in the sub-regions dominated by a single elongated structure, or ``ridge'', and the sub-regions dominated by multiple filamentary structures, or ``nests''.
Towards the Centre-Ridge, where RCW\,36, the highest maximum values of \nh, the shallowest \nh-PDF power-law tail, and most of the massive cores are found, we see that the iso-\nh\ contours are mostly perpendicular to \bperp.
Towards the South Ridge sub-region, where the slope of the \nh-PDF power-law tail is comparable to the Centre-Ridge sub-region, we see a steep transition from \nh\ structures and \bperp\ being mostly parallel at \lognh\,$<22.4$ to mostly perpendicular at \lognh\,$>22.4$.}
\juan{In contrast, towards the Centre-Nest and the South-Nest, where the power-law tails of the \nh-PDF are steeper and the maximum \nh\ values are lower than in the ridge-like regions, we see a less steep transition from \nh\ structures and \bperp\ being mostly parallel at low \nh\ to mostly perpendicular at the highest \nh.}
\juan{We found a similar behaviour towards the Serpens South region in Aquila, where we compared the column density structure inferred from the \Herschel\ satellite observations and \bperp\ inferred from observations of polarization by \Planck\ at 353\,GHz (850\micron).}
These observational results suggest that the magnetic field is important in gathering the gas that composes the molecular cloud, and \juan{a} remnant of that assembly process is present in the relative orientation of the projected density and magnetic field.

\juan{Future studies of the molecular emission towards Vela\,C would enable the understanding of the kinematics in this region, which would help determine how the ridge-like and the nest-like sub-regions came to be.
Further comparison of the relative orientation between the \nh\ structures and \bperp\ with the population of cores in other molecular clouds would also help us understand how the magnetic field influences the formation of density structures, and potentially how the cloud-scale magnetic environment is related to the formation of stars.}

\begin{acknowledgements}
The BLASTPol collaboration acknowledges support from NASA through grant numbers NAG5-12785, NAG5-13301, NNGO-6GI11G, NNX0-9AB98G, and the Illinois Space Grant Consortium, the Canadian Space Agency, the Leverhulme Trust through the Research Project Grant F/00 407/BN, Canada's Natural Sciences and Engineering Research Council, the Canada Foundation for Innovation, the Ontario Innovation Trust, and the US National Science Foundation Office of Polar Programs.
This work was possible through the funding from the European Research Council under the European Community's Seventh Framework Programme (FP7/2007-2013 Grant Agreement no. 306483 and no. 291294).
We are grateful to Davide Elia and Theresa Giannini for providing their unpublished catalog\juan{ue} of dense cores in Vela\,C. 
L.~M.~F. is a Jansky Fellow of the National Radio Astronomy Observatory (NRAO). NRAO is a facility of the National Science Foundation (NSF operated under cooperative agreement by Associated Universities, Inc.
F.~P. thanks the European Commission under the Marie Sklodowska-Curie Actions within the H2020 program, Grant Agreement number: 658499 PolAME H2020-MSCA-IF-2014.
We thank the Columbia Scientific Balloon Facility staff for their outstanding work.
\end{acknowledgements}

\bibliographystyle{aa}
\bibliography{BLASTPolHRO.bbl}

% -------------------------------------------------------------------------------------------------------------------------------------------------------------------------
% APPENDICES APPENDICES APPENDICES APPENDICES APPENDICES APPENDICES APPENDICES APPENDICES -------------------------------------------------------------------------------------------------------------------------------------------------------------------------
\appendix

\section{The $I_{500}$ HROs}\label{appendix:HROI}

We construct the \juan{histograms of the relative orientations (HROs)} using BLASTPol observations of polarization at 250, 350, and 500\micron\ and the gradient of the intensity observed by BLASTPol in the $500$\micron\ band, $I_{500}$, using the procedure described in Sect.~\ref{section:hroconstruction}.
This family of HROs has the advantage of considering observations made with the same instrument, but the obvious disadvantage that $I_{500}$ is merely a proxy for the total gas column density.

\begin{figure*}[ht!]
\centerline{
\includegraphics[width=0.33\textwidth,angle=0,origin=c]{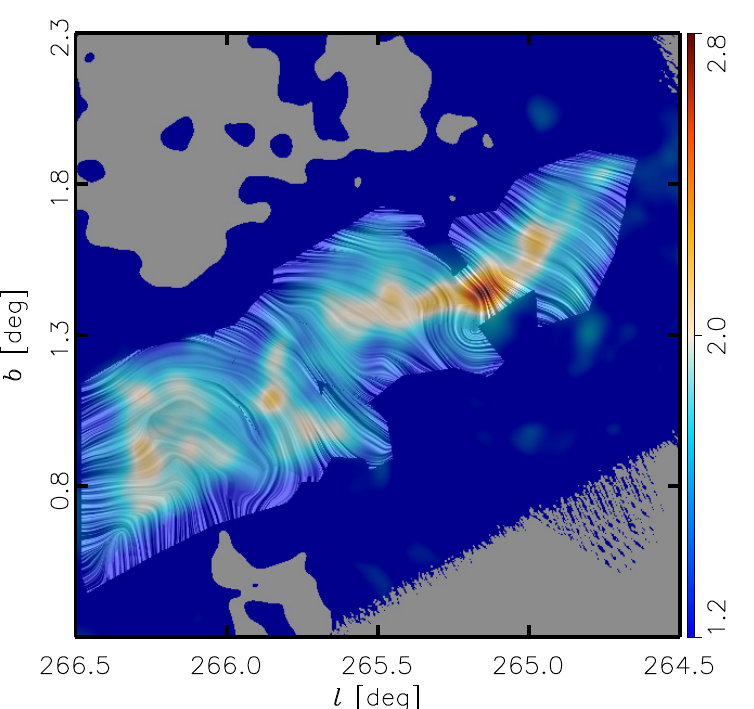}
\includegraphics[width=0.33\textwidth,angle=0,origin=c]{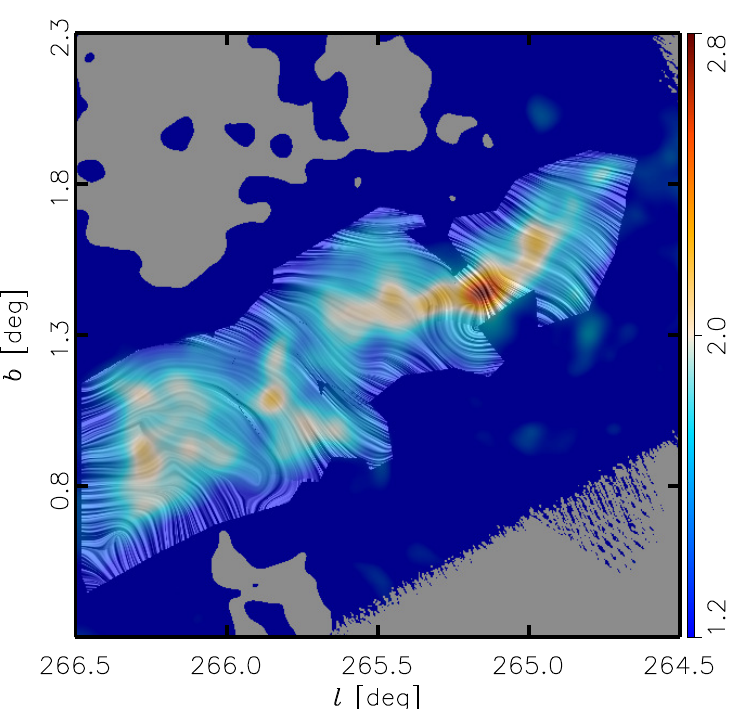}
\includegraphics[width=0.33\textwidth,angle=0,origin=c]{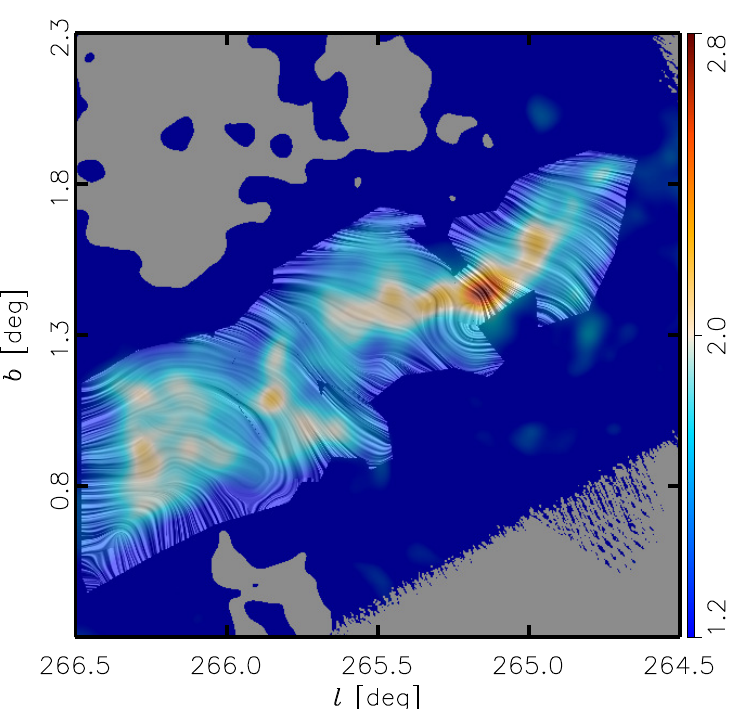}
}
\vspace{-1.0mm}
\centerline{
%\hspace{-0.01\textwidth}\includegraphics[width=0.32\textwidth,angle=0,origin=c]{HROIhist250micronAll_int.eps}
%\hspace{0.01\textwidth}\includegraphics[width=0.32\textwidth,angle=0,origin=c]{HROIhist350micronAll_int.eps}
%\hspace{0.01\textwidth}\includegraphics[width=0.32\textwidth,angle=0,origin=c]{HROIhist500micronAll_int.eps}
\hspace{-0.01\textwidth}\includegraphics[width=0.32\textwidth,angle=0,origin=c]{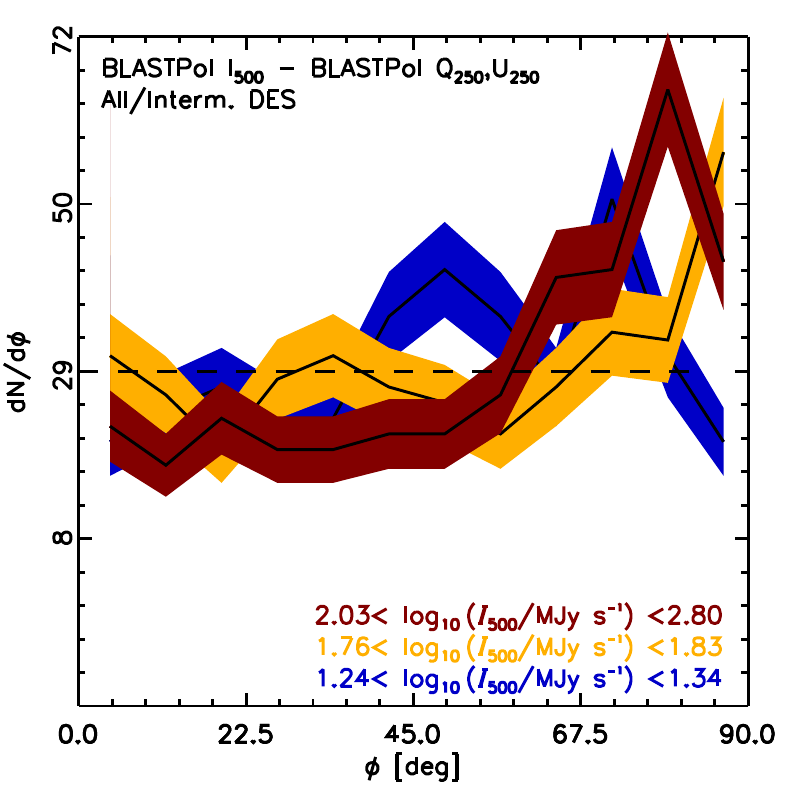}
\hspace{0.01\textwidth}\includegraphics[width=0.32\textwidth,angle=0,origin=c]{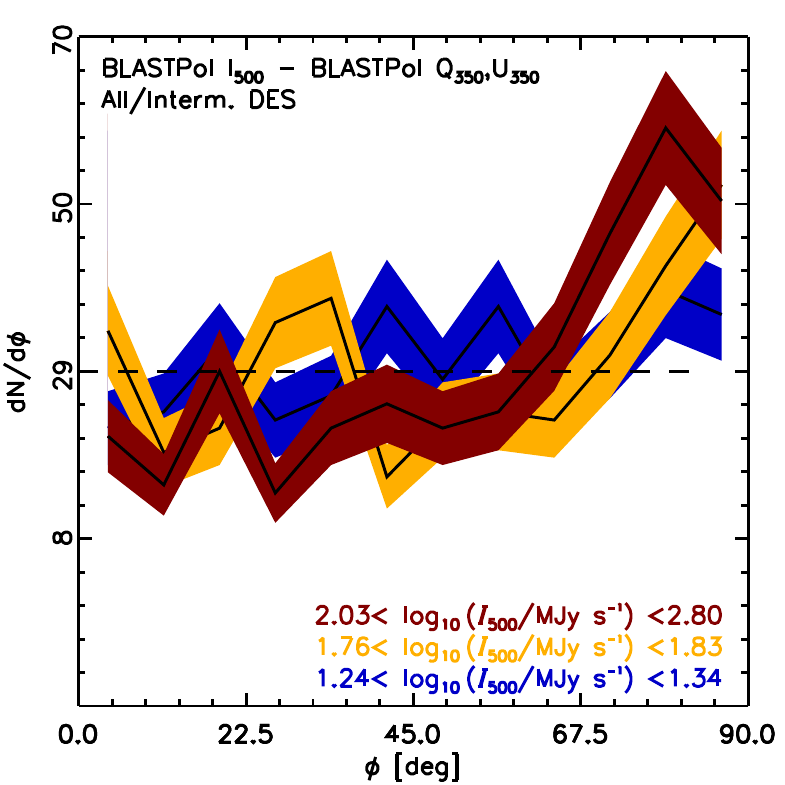}
\hspace{0.01\textwidth}\includegraphics[width=0.32\textwidth,angle=0,origin=c]{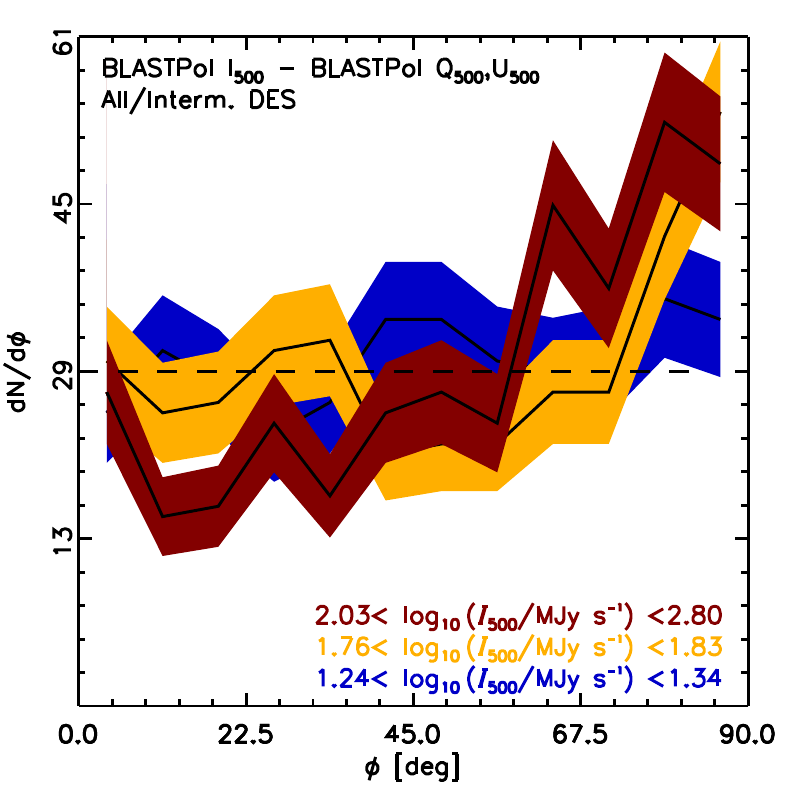}
}
\caption{\emph{Top}. Magnetic field and total intensity measured by BLASTPol towards two regions in the Vela molecular ridge. 
The colours represent $I_{500}$, the total intensity observed in the 500\juan{-$\mu$m} wavelength-band of BLASTPol. 
The ``drapery'' pattern, produced using the line integral convolution method \citep[LIC,][]{cabral1993}, indicates the orientation of magnetic field lines, orthogonal to the orientation of the submillimetre polarization observed by BLASTPol at 250 (\emph{left}), 350 (\emph{centre}), and 500\micron\ (\emph{right}).
\emph{Bottom}. 
Histogram of the relative orientations (HRO) between the iso-$I_{500}$ contours and the magnetic field orientation inferred from the BLASTPol observations at 250 (\emph{left}), 350 (\emph{centre}), and 500\micron\ (\emph{right}).
The figures present the HROs for the lowest, an intermediate, and the highest $I_{500}$ bin (blue, orange, and dark red, respectively). 
The bins have equal numbers of selected pixels within the $I_{500}$-ranges labelled.
The horizontal dashed line corresponds to the average per angle bin of 15\deg.
The widths of the shaded areas for each histogram correspond to the 1-$\sigma$ uncertainties related to the histogram binning operation. 
Histograms peaking at 0\deg\ correspond to \bperp\ predominantly aligned with iso-$I_{500}$ contours. 
Histograms peaking at 90\deg\ correspond to \bperp\ predominantly perpendicular to iso-$I_{500}$ contours.
}
\label{fig:HROblastpolblastpol}
\end{figure*}

\begin{figure}[ht!]
\centerline{
\includegraphics[width=0.5\textwidth,angle=0,origin=c]{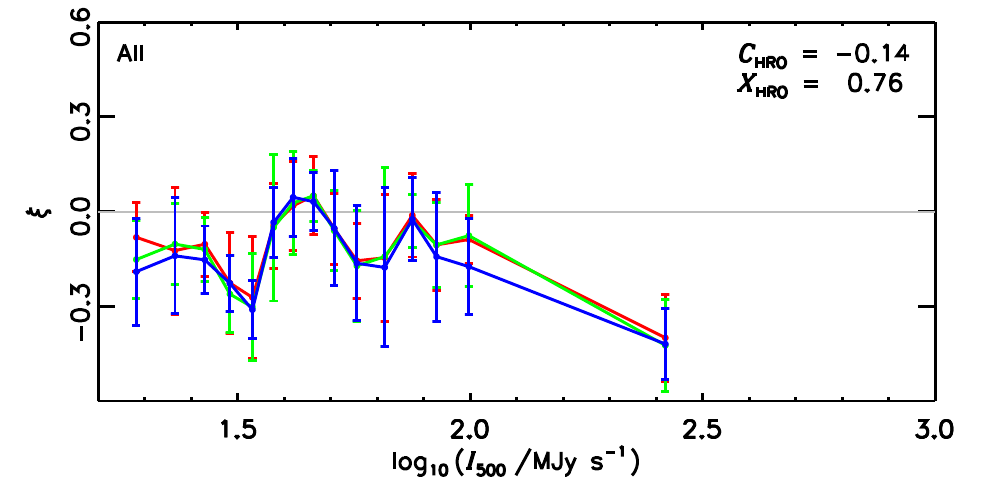}
}
\vspace{-1.0mm}
\caption{Relative orientation parameter $\xi$, defined in Eq.~\ref{eq:zeta}, calculated for the different $I_{500}$ bins towards the Vela\,C region. 
The values $\xi > 0$ and $\xi < 0$ correspond to the magnetic field oriented mostly parallel and perpendicular to the iso-$I_{500}$ or iso-\nh\ contours, respectively. 
The grey line corresponds to $\xi=0$, which corresponds to the case there is not a preferred relative orientation.
}
\label{fig:hrozetaI}
\end{figure}

The $I_{500}$ HROs, shown in the lower panels of Fig.~\ref{fig:HROblastpolblastpol}, are flat in the lowest $I_{500}$ \juan{range} and peak at 90\deg\ in the intermediate and highest $I_{500}$ \juan{ranges} across the three BLASTPol wavelength bands.
The agreement between observations in different \juan{wavebands} and the prevalence of the HROs peaking at 90\deg\ is confirmed by the behaviour of $\xi$ as a function of $I_{500}$, shown in Fig.~\ref{fig:hrozetaI}.
The figure shows that $\xi<0$ in almost all of the $I_{500}$ bins, although the error bars indicate that this value is only clearly separated from $\xi\approx0$ in the highest $I_{500}$ bin.

However, the $I_{500}$ HROs \juan{that} indicat\juan{e} that \bperp\ is mostly perpendicular to $I_{500}$ are not conclusive given that $I_{500}$ is only a proxy \juan{to} column density.
Direct comparison with the trends presented in \cite{planck2015-XXXV} would require further evaluation of the total gas column density, something that is readily available through the multi-wavelength observation of this region by \Herschel.

% -------------------------------------------------------------------------------------------------------------------------------------------------------------------------
\section{Diffuse emission subtraction}\label{appendix:refRegions}

\begin{figure*}[ht!]
\centerline{
\includegraphics[width=0.32\textwidth,angle=0,origin=c]{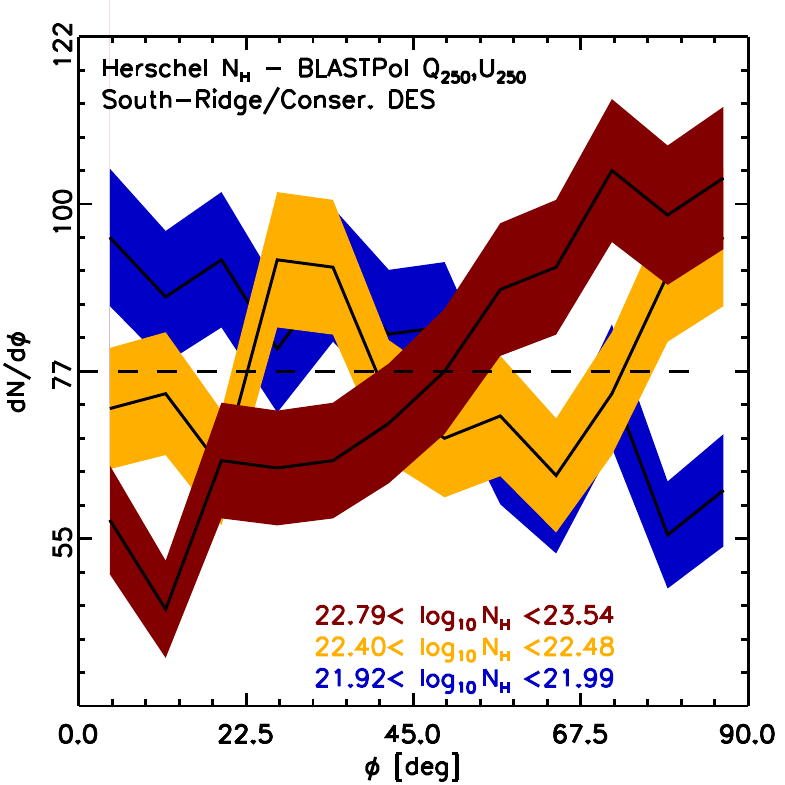}
\includegraphics[width=0.32\textwidth,angle=0,origin=c]{HROhist250micronSouth-Ridge_int-eps-converted-to.pdf}
\includegraphics[width=0.32\textwidth,angle=0,origin=c]{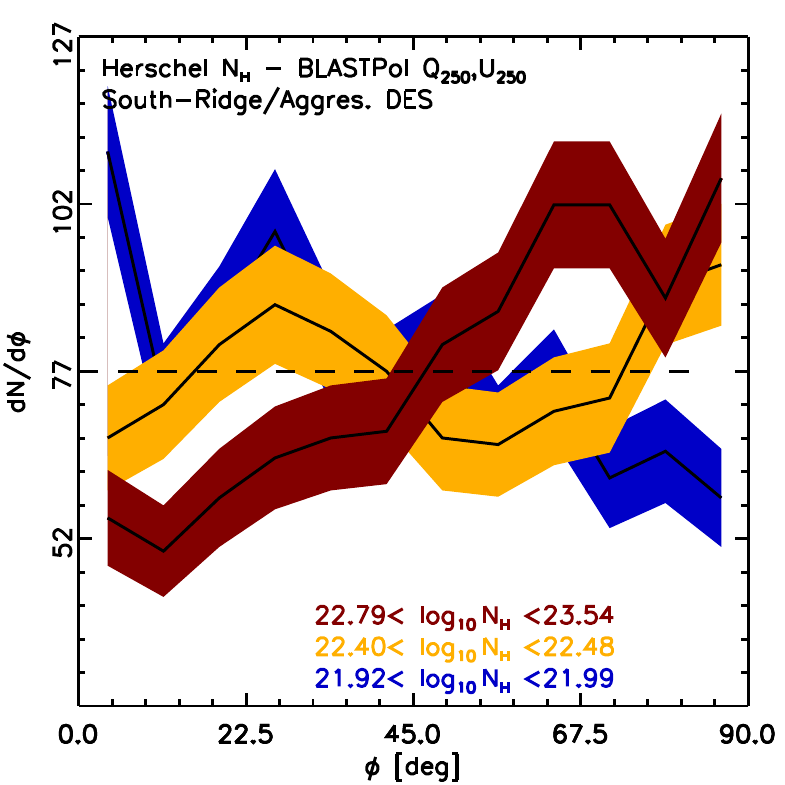}
}
\caption{Histogram of the relative orientations (HRO) between the iso-\nh\ contours and the magnetic field orientation inferred from the BLASTPol observations at 250\micron\ obtained with the three diffuse emission subtraction (DES) methods, \juan{namely} aggressive, intermediate, and conservative\juan{,} defined in Section~\ref{subsection:BLASTPol}.
The figures present the HROs for the lowest \juan{bin}, an intermediate \juan{bin}, and the highest \nh\ bin (blue, orange, and dark red, respectively). 
The bins have equal numbers of selected pixels within the \nh\ ranges labelled.
The horizontal dashed line corresponds to the average.
The widths of the shaded areas for each histogram correspond to the 1-$\sigma$ uncertainties related to the histogram binning operation. 
Histograms peaking at 0\deg\ \juan{would} correspond to \bperp\ \juan{being} predominantly aligned with iso-\nh\ contours\juan{, while} histograms peaking at 90\deg\ \juan{would} correspond to \bperp\ predominantly perpendicular to iso-\nh\ contours.
}
\label{fig:HRO250DEStest}
\end{figure*}

The observed polarization degree and orientation angles are the product of the contributions from aligned dust grains in the cloud and in the foreground and background. 
In order to estimate the contribution of the foreground/background, \cite{fissel2016} present three different background subtraction methods, one conservative and one more aggressive with respect to diffuse emission subtraction (DES), as described in Sect.~\ref{subsection:BLASTPol}.

\begin{figure}[ht!]
\centerline{
\includegraphics[width=0.5\textwidth,angle=0,origin=c]{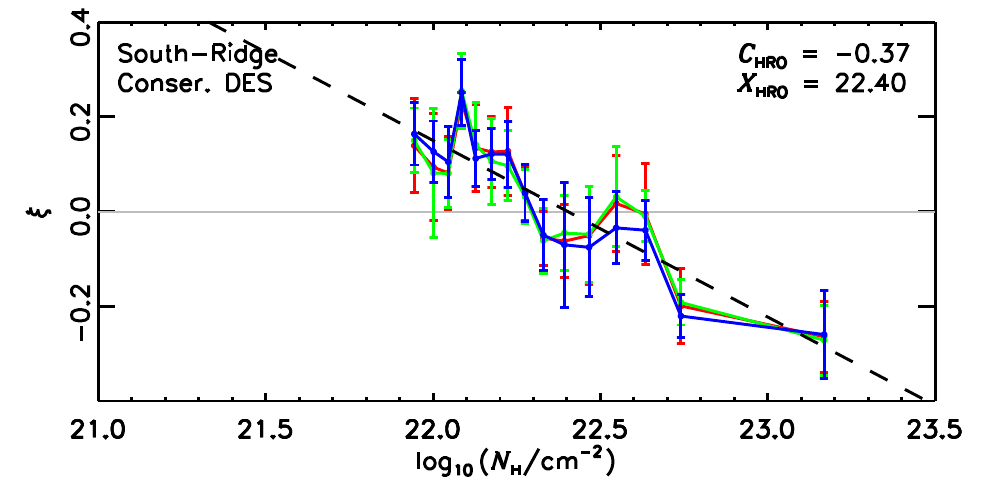}
}
\vspace{-1.5mm}
\centerline{
\includegraphics[width=0.5\textwidth,angle=0,origin=c]{VelaCSouth-RidgeREFint_HROzeta-eps-converted-to.pdf}
}
\vspace{-1.5mm}
\centerline{
\includegraphics[width=0.5\textwidth,angle=0,origin=c]{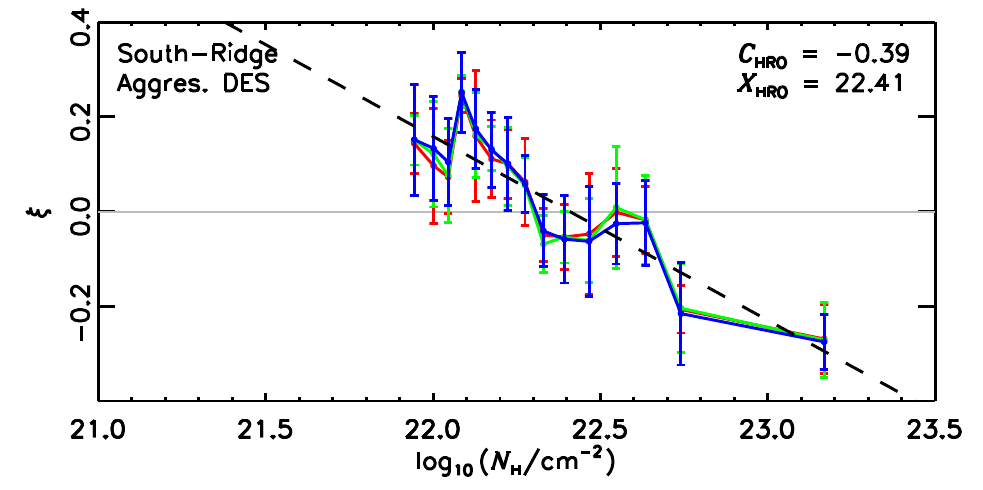}
}
\vspace{-1.0mm}
\caption{Relative orientation parameter $\xi$, defined in Eq.~\ref{eq:zeta}, calculated for the different \nh\ bins towards the two sub-regions of Vela\,C corresponding to the three diffuse emission subtraction (DES) methods; aggressive, intermediate, and conservative; defined in Section~\ref{subsection:BLASTPol}.
The values $\xi > 0$ and $\xi < 0$ correspond to the magnetic field oriented mostly parallel and perpendicular to the iso-\nh\ contours, respectively. 
\juan{The black dashed line and the values of $C_{\textsc{HRO}}$ and $X_{\textsc{HRO}}$ correspond to the linear fit introduced in Eq.~\ref{eq:hrofit}}.
The grey line is $\xi=0$, which corresponds to the case \juan{where} there is no preferred relative orientation.
}
\label{fig:zetaDEStest}
\end{figure}

Figure~\ref{fig:HRO250DEStest} shows the HROs calculated using the BLASTPol 250\juan{-}$\mu$m observations\juan{,} estimated using the three DES methods\juan{,} towards the South-Ridge sub-region of Vela\,C.
The main differences between the HROs corresponding to the different DES methods are the \juan{amount of jitter}, which \juan{is} similar in the conservative and intermediate DES, but slightly different in the HROs that correspond to the aggressive DES.
However, the HROs in Fig.~\ref{fig:HRO250DEStest} indicate that the selection of the reference regions for the DES do\juan{es} not significantly affect the relative orientation trends that they describe.
This agreement between the HROs corresponding to different DES methods is also found in the other sub-regions of Vela\,C.
The values of the relative orientation parameter, $\xi$, shown in Fig.~\ref{fig:zetaDEStest}, also reveal that the selection of DES \juan{method} does not significantly change the behaviour of $\xi$ as a function of \nh. 

% ============================================================================================
\section{The Aquila region}\label{appendix:AquilaRegion}

\begin{figure}[ht!]
\vspace{-1.0mm}
\centerline{
\includegraphics[width=0.48\textwidth,angle=0,origin=c]{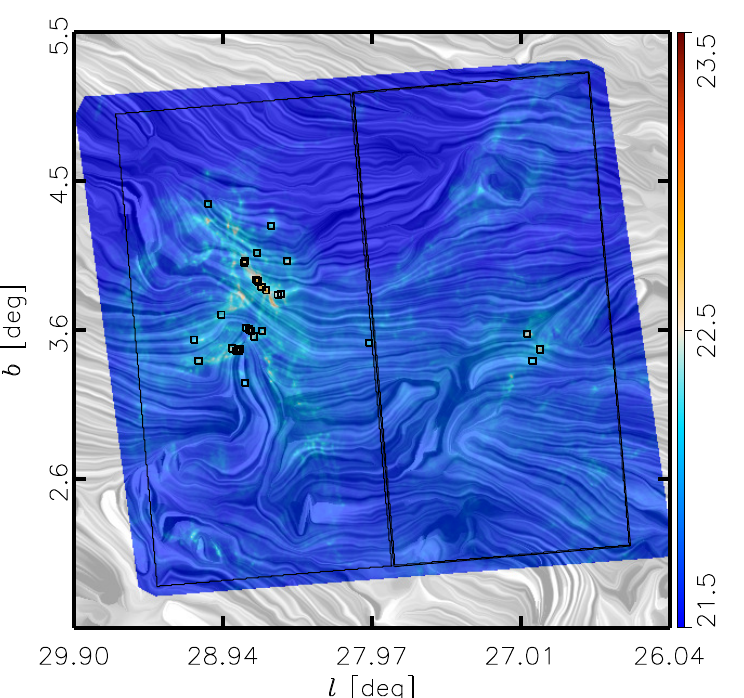}
}
\vspace{-1.0mm}
\caption{
Magnetic field and total gas column density measured towards a sub-region of the Aquila rift. 
The colours represent the total gas column density, \nh, inferred from the \Herschel\ observations \citep{konyves2015}. 
The ``drapery'' pattern, produced using the line integral convolution method \citep[LIC,][]{cabral1993}, indicates the orientation of magnetic field lines, orthogonal to the orientation of the submillimetre polarization observed by \Planck\ at 353\GHz.
The black squares correspond to the location of the cores with $M>3$\,M$_{\odot}$\juan{, pre-stellar and protostellar,} from the catalog\juan{ue} presented in \cite{konyves2015}.
The black polygons correspond to the two portions where we compare the results of the HRO analysis, which contain W40 (\juan{left}) and MWC\,297 (\juan{right}), respectively.
}
\label{fig:AquilaLIC}
\end{figure}

\begin{figure}[ht!]
\centerline{
\includegraphics[width=0.49\textwidth,angle=0,origin=c]{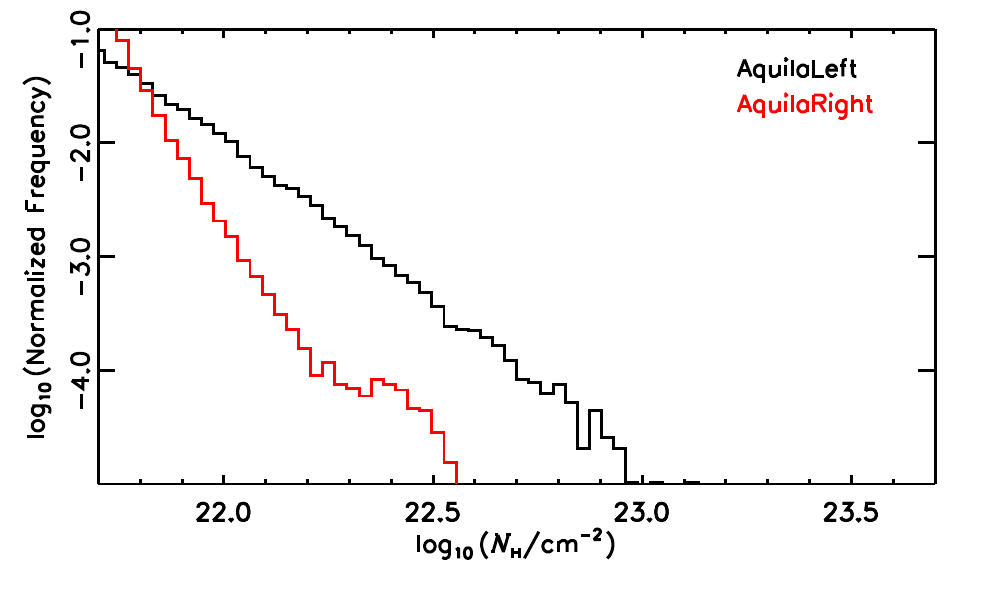}
}
\vspace{-3.0mm}
\caption{\juan{Comparison of the total gas column density probability distribution functions (PDFs), inferred from the \Herschel\ observations \citep{konyves2015}, in the two sub-regions of the Aquila rift presented in Fig.~\ref{fig:AquilaLIC}.
The y-axis is presented in logarithmic scale for the sake of comparison with figure~5 of \cite{hill2011}.}
}
\label{fig:AquilaNHPDF}
\end{figure}

\begin{figure}[ht!]
\centerline{
\includegraphics[width=0.4\textwidth,angle=0,origin=c]{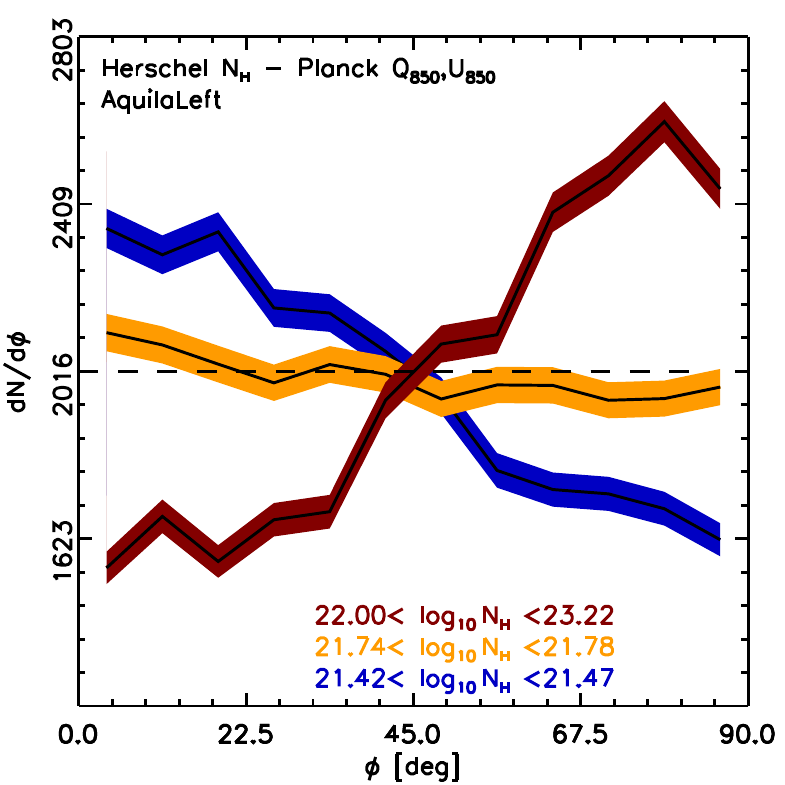}
}
\centerline{
\includegraphics[width=0.4\textwidth,angle=0,origin=c]{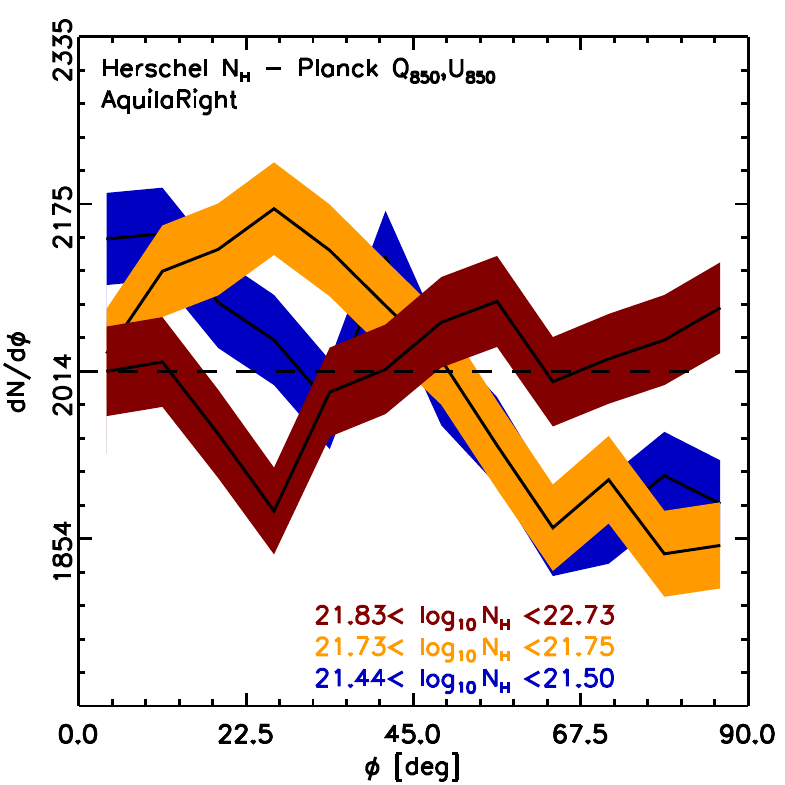}
}
\caption{
Histogram of the relative orientations (HRO) between the iso-$N_{\rm H_{2}}$ contours and the magnetic field orientation inferred from the \Planck\ 353\,GHz observations towards the two regions of the Aquila rift illustrated in Fig.~\ref{fig:AquilaLIC}.
The figures present the HROs for the lowest, an intermediate, and the highest \nh\ bin (blue, orange, and dark red, respectively). 
The bins have equal numbers of selected pixels within the \nh-bins ranges labelled.
The horizontal dashed line corresponds to the average per angle bin of 15\deg.
The widths of the shaded areas for each histogram correspond to the 1-$\sigma$ uncertainties related to the histogram binning operation. 
Histograms peaking at 0\deg\ correspond to \bperp\ predominantly aligned with iso-$N_{\rm H_{2}}$ contours. 
Histograms peaking at 90\deg\ correspond to \bperp\ predominantly perpendicular to iso-$N_{\rm H_{2}}$ contours.
}
\label{fig:AquilaExperimentHRO}
\end{figure}
\begin{figure}[ht!]
\centerline{
\includegraphics[width=0.5\textwidth,angle=0,origin=c]{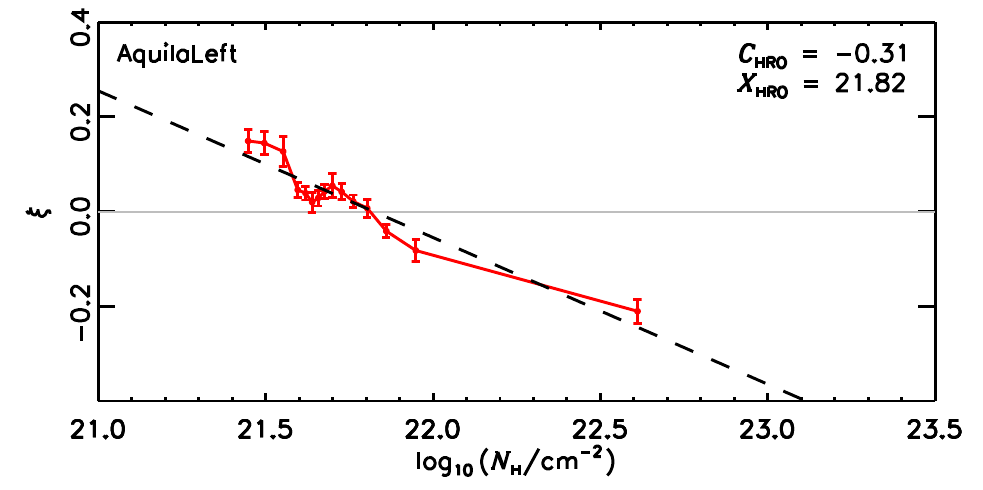}
}
\centerline{
\includegraphics[width=0.5\textwidth,angle=0,origin=c]{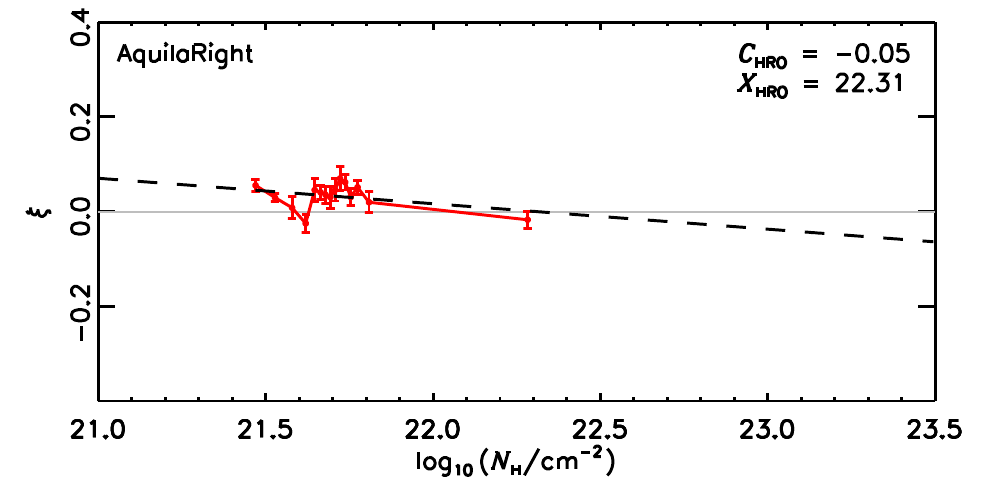}
}
\vspace{-1.0mm}
\caption{Relative orientation parameter $\xi$, defined in Eq.~\ref{eq:zeta}, calculated for the different \nh\ bins towards the two portions of the Aquila rift illustrated in Fig.~\ref{fig:AquilaLIC}. 
The values $\xi > 0$ and $\xi < 0$ correspond to the magnetic field oriented mostly parallel and perpendicular to the iso-\nh\ contours, respectively. 
\juan{The black dashed line and the values of $C_{\textsc{HRO}}$ and $X_{\textsc{HRO}}$ correspond to the linear fit introduced in Eq.~\ref{eq:hrofit}}.
The grey line is $\xi=0$, which corresponds to the case \juan{where} there is no preferred relative orientation.
}
\label{fig:AquilaExperimentZeta}
\end{figure}

The Aquila rift is \juan{a} 5\deg-long extinction feature above the Galactic plane at \juan{G}alactic longitudes between $l=30$\deg\ and $l=50$\deg\ \citep{prato2008}.
In this work, we focus on the 3\deg\,$\times$\,3\deg\ portion of Aquila rift around $l=28$\deg\ and $b=3\pdeg5$ observed by \Herschel\ SPIRE at 250, 350,  and 500\micron\ and PACS at 70 and 160\micron\ \citep{bontemps2010}.
This region observed by \Herschel\ includes: Serpens South, a young protostellar cluster showing very active recent star formation and embedded in a dense filamentary cloud; W40, a young star cluster associated with the eponymous \ion{H}{ii} region; and MWC\,297, a young 10\,M$_\odot$ star \citep[][and the references therein]{konyves2015}.

The whole area covered by \Herschel\ has a total mass of $3.1\,\times\,10^4$\,M$_\odot$, with a $1.1\,\times\,10^4$\,M$_\odot$ region associated with W40, and a $4.1\,\times\,10^3$\,M$_\odot$ region associated with MWC\,297 \citep[][and references therein]{bontemps2010}.
Although this portion of Aquila is less massive than Vela\,C, observations with P-ArT\'{e}MiS indicate that the filamentary structures in both regions are similar in column density profiles and mass per unit length \citep{hill2012}, making it an interesting candidate for comparison using the HRO analysis.

%In terms of mass, the MC studied in \cite{planck2015-XXXV} that is the closest to Vela\,C is Orion \citep{bally2008}, but its direct comparison in terms of HRO\juan{s} is not straightforward.
%On the one hand, Orion is located at roughly twice the distance to Serpens, making the gap in resolution between the \Planck\ 353-GHz polarization and the \Herschel\ \nh\ maps even larger than in our study of Vela\,C.
%On the other hand, Orion is composed of multiple regions located at a variety of distances, with a complex recent star-formation history that complicates the interpretation of the HRO analysis.
%This does not mean that the study of the relation between the relative orientations between \nh\ and \bperp\ should not be extended to the Orion region, but simply that such \juan{a} study is beyond the scope of this paper, where we focus on just one simple example towards a relatively more quiescent region.

\subsection{Observations}

Fig.~\ref{fig:AquilaLIC} shows the \nh\ and \bperp\ maps, 
%derived from the \planck\ 353\juan{-}GHz polarization observations
and the position of the clumps with $M>3$\,M$_{\odot}$ from the catalog\juan{ue} of dense cores presented in \cite{konyves2015}.
At the assumed distance \juan{of} the Aquila rift, 260\,pc, these maps are sampling physical scales of 0.12 and 2\,pc, respectively. 

\subsubsection{Thermal dust polarization}

We use the \Planck\ 353\,GHz Stokes $Q$ and $U$ maps and the associated noise maps made from five independent consecutive sky surveys of the \Planck\ cryogenic mission, which together correspond to the \Planck\ 2015 public data release\footnote{\url{http://pla.esac.esa.int/pla/}}~\citep{planck2014-a01}. 
The whole-sky 353\,GHz maps of $Q$ and $U$, their respective variances $\sigma^{2}_{\textsc{Q}}$ and $\sigma^{2}_{\textsc{U}}$, and their covariance $\sigma_{\textsc{QU}}$ are initially at 4\parcm8 resolution in \healpix\ format\footnote{\url{http://healpix.sf.net}} \citep{gorski2005}\juan{,} with a pixelization at $N_{\rm side} = 2048$, which corresponds to an effective pixel size of 1\parcm7. 
To increase the signal-to-noise ratio (S/N) of extended emission, we smooth all the maps to 10\arcmin\ resolution using a Gaussian approximation to the \Planck\ beam and the smoothing procedures for the covariance matrix described in \cite{planck2014-XIX}.

The maps of the individual regions are projected and resampled onto a Cartesian grid with the gnomonic projection procedure described in \cite{paradis2012}. The present analysis is performed on these projected maps. 
The selected regions are small enough, and are located at sufficiently low Galactic latitudes\juan{,} that this projection does not impact significantly on our study.

\subsubsection{Column density}

We use the 36\parcs5 resolution $N_{\rm H_{2}}$ column density maps derived from the 70\juan{-}, 160\juan{-}, 250\juan{-}, 350\juan{-}, and 500\juan{-$\mu$m} \Herschel\ observations, described in \cite{konyves2015} and publicly available in \juan{the} archive of the \Herschel\ Gould Belt Survey\footnote{\url{http://www.herschel.fr/cea/gouldbelt}}~\citep[HGBS,][]{andre2010} \juan{project}.

Additionally, we use the catalogue of dense cores identified in the HGBS maps of the Aquila region \citep{konyves2015}, considering their estimated core mass obtained by assuming the dust opacity law advocated by \cite{roy2014}.

\subsection{Analysis}

We appl\juan{y} the analysis described in Sect.~\ref{section:analysis}\juan{,} using the aforementioned maps and focus on two arbitrary sub-regions with the same area: one containing the W40 \ion{H}{ii} region; and \juan{the other with} MWC\,297.
Given, that the region around W40 contains the majority of the candidate prestellar and protostellar cores, as clearly show in figure~1 of \cite{konyves2015}, \juan{and a shallower high-\nh\ PDF tail, as shown in Fig.~\ref{fig:AquilaNHPDF},} we aim to evaluate if \juan{both characteristics are} correlated with the behaviour of the HROs.

Figure~\ref{fig:AquilaExperimentHRO} shows the HROs of the two sub-regions of the Aquila rift. 
The region containing W40, where most of the candidate prestellar and protostellar cores are located, shows a clear change from the histogram peaking at 0\deg\ in the lowest $N_{\rm H_{2}}$-bin \juan{(}indicating \bperp\ mostly parallel to the \nh\ structures\juan{)} to the histogram peaking at 90\deg\ in the highest \nh-bin \juan{(}indicating \bperp\ mostly perpendicular to the \nh\ structures\juan{)}.
In contrast, the region containing MWC\,297, shows histograms peaking at 0\deg\ in the lowest and intermediate \nh-bins, while the HRO in the highest \nh-bin has a lot of jitter, but is consistent with no preferential relative orientation between \nh\ and \bperp.
However, the region containing MWC\,297 has a lower maximum \nh\ value, \juan{and} thus, a more complete evaluation of the changes in the HRO should be made in terms of $\xi$.

Fig.~\ref{fig:AquilaExperimentZeta} shows the \juan{relative orientation} parameter, $\xi$ as defined in Eq.~\ref{eq:zeta}, as a function of \nh\ in both sub-regions of the Aquila rift.
The values of $\xi$ and $C_{\rm HRO}$ indicate \juan{that} the region containing W40 has a clear change in the relative orientation between \nh\ and \bperp, from mostly parallel to mostly perpendicular with increasing \nh.
In the region containing MWC\,297, $C_{\rm HRO}$ is \juan{also} closer to zero and $\xi$ is closer to zero, suggesting a small change in relative orientation between \nh\ and \bperp\ from mostly parallel to mostly perpendicular with increasing \nh\, but \juan{with} values of $\xi$ suggesting no preferential relative orientation.

The behaviour of $\xi$ as a function of \nh\ is in clear agreement with that reported for a much larger portion of the Aquila rift in \cite{planck2015-XXXV}; \juan{however,} the comparison with the higher-resolution \Herschel\ map shows that different portions of the cloud have different degrees of change in relative orientation between \nh\ and \bperp.
In the sub-region of the Aquila rift presented in this work, we found a sharper transition between \nh\ being preferentially parallel or having no preferred relative orientation with respect to \bperp\ at the low \nh, to being preferentially perpendicular at the highest \nh, in the portion of the cloud with the greatest values of \nh, the flattest high-column density tail of the PDF, and the largest number of prestellar and protostellar cores.

\raggedright

\end{document}